\documentclass[12pt,preprint,prb,floatfix]{revtex4-1}  

 \usepackage{natbib}
 \usepackage{graphicx}
\usepackage{subcaption}
 \usepackage{dcolumn}
 \usepackage{amsmath}
 \usepackage{textcomp}
 \usepackage{float}
 \usepackage{rotating}
 \usepackage{stackrel}
 \usepackage{hyperref}

  \DeclareMathAlphabet{\mathds}{OT1}{pzc}{m}{it}
 
\begin{document}

 \bibliographystyle{prsty} 
 
 \title{Theory of Lineshapes in Optical-Optical Double Resonance Spectroscopy}
 
 \author{Kevin K. Lehmann}
 \address{
 Departments of Chemistry \& Physics, University of Virginia, Charlottesville VA, 22904-4319}
 \date{\today}

\begin{abstract}
This paper presents a theory for  lineshapes for molecular Optical-Optical Double Resonance (DR) Spectroscopy with arbitrary strength for both pump and probe field using the steady-state solutions for the 3-level density matrix.   When the Doppler broadening can be neglected, the results are analytical, and the probe spectrum is a pair of Lorentzian lines that display Autler-Townes splitting and each has angular frequency half-width half maximum equal to the relaxation rates, which are all assumed equal.   When Doppler broadening is introduced, one must resort to numerical integration except for the limit of weak pump and probe fields.   When the Doppler width is assumed much larger than the pump and probe Rabi Frequencies and the probe frequency higher than that of the pump,, the calculated DR lineshapes are found to be Lorentzian with a strong pump field limit that is proportional to the pump Rabi frequency, what is commonly known as power broadening.   However, the width does not equal that Rabi frequency and is different for co- and counter-propagating pump and probe fields.  Furthermore, that broadening is largely inhomogeneous, despite the Lorentzian shape.  The saturation power is found to be about 4 times higher than for the bare probe transition with the same relaxation rate, dramatically lower than that expected if the width is interpreted as homogeneous.
\end{abstract}
 \maketitle

 Double-Resonance (DR) is one of the most powerful experimental methods in the spectroscopist's tool belt.\cite{DemtroderLS2}  
 Originally used for RF and Microwave transitions,~\cite{Autler55} due to the early development of coherent radiation sources
 in those frequency domains, the development of optical-optical double resonance (OODR) had to await the development of lasers. 
 Due to the very limited tuning range of early lasers, initial applications were to atomic laser transitions.~\cite{Putlitz65, Series66a}
 Molecular RF-IR and $\mu$W-IR DR was observed using coincidences of molecular ro-vibrational transitions with IR lasers, such
 as the 3.3\,$\mu$m He-Ne laser transition that overlapped P(7) transitions of the CH$_4$ $\nu_3$ fundamental~\cite{Curl73b,Curl73c}
while (RF,$\mu$W)-Vis DR was observed using overlap of ion laser lines with molecular electronic transitions.~\cite{Field73, Solarz74}  
Molecular OODR became practicable with the
development of broadly tunable dye lasers.~\cite{Field75}

There is a rich literature of theoretical treatments for the interaction of three quantum levels with two radiation fields, a situation that
applies to most DR experiments.  Influential early papers include work by Feld and Javan\cite{Feld69} and H{\"a}nsch and Toschek.~\cite{Hansch70}
 A particularly lucid treatment can be
found in the text by Berman and Malinowsky.~\cite{Berman11}  Most commonly, these treatments evaluate the steady-state solutions
of the Optical-Bloch equations.~\cite{Scully02}   Most such treatments have modeled atomic spectroscopy for which spontaneous 
emission is a dominant relaxation mechanism and there is a strong state dependence to even the collisional relaxation rates.  
These atomic models typically are for closed sets of levels with explicit emission rates between states in the model.  For
molecular problems, given the large number of states in the system, one treats the resonant levels as part of an open system
coupled to a large bath of levels, and one can usually assume that the populations of the bath states are negligibly perturbed by
the optical fields, though that must be relaxed to model collision induced transitions, so called four level DR,\cite{Oka66} where the
disequilibrium produced by the pump wave is collisionally transferred to one or both of the states involved in the probe transition.
Recently, the
present author and collaborators have been using IR-IR DR resonance spectroscopy to study vibrational spectroscopy of methane.\cite{Foltynowicz21a, Foltynowicz21b,
Lehmann23, deOliveira24, Hjalten24, Lehmann25, deOliveira25}  A single-mode Optical Parametric Oscillator (OPO) was used to pump selected transitions in the $\nu_3$ fundamental CH$_4$
and a near-IR frequency comb was used to probe the sub-Doppler width absorption in the spectral region of the $3\nu_3 \leftarrow \nu_3$ transitions. 
Similar IR-IR DR spectra have been observed on other molecules using  pairs of cw lasers.~\cite{Karhu17, Hu21, Okubo21, Tan24}
This paper presents the theoretical framework we have developed to model these experiments.  While closely related to prior work on
three level systems, the results are considerably simplified by the assumption that relaxation rates of both populations and coherences are
all equal.  We believe this is an excellent approximation for such ro-vibrational DR experiments with monochromatic pump lasers
due to the fact that for such transitions the spontaneous
emission rates are negligible compared to collisional relaxation (or even diffusion out of the laser beams) and that the collisional relaxation is
dominated by the total scattering cross section, which has very small variation between different ro-vibrational states, being dominated by the long range
interaction potentials. 

In OODR experiments where the radiation fields were monochromatic, the resulting spectral features are much narrower than thermal Doppler widths and often described as
 ``Doppler Free''.  However, the Bennett hole and hill produced by the pump laser have a finite spread of velocity component along
 the propagation direction and this results in a Doppler contribution to the width of DR transitions.\cite{Bennett62}  At low powers for the pump and
 probe fields (Rabi frequencies significantly less than the relaxation rates) the DR lines have half widths are half maximum (HWHM) proportional to the
 homogeneous relaxation rates for the transitions involved. 
 At higher pump power, the DR spectra are broadened and display what
 is commonly called ``power broadening'' with HWHM proportional to, but not equal to, the pump Rabi frequency.
 When Doppler broadening of pump and probe transitions can be neglected, simple analytical results for the positions
 and widths of the DR features are found, consistent with earlier work, but simplified by the assumption of a single
 relaxation rate.   Modeling of thermal samples requires convolution of the homogeneous spectrum over the distribution of
 Doppler shifts of the pump and probe transitions.  As in previous work,\cite{Berman11} this requires numerical integration to evaluate
 the DR lineshape.  In the limit of a Doppler width much greater than the Rabi frequencies and relaxation rates, the
 lineshapes are found to be Lorentzian when the wavenumber of probe is greater than that of the probe.
 When the pump Rabi frequency greatly exceeds the relaxation rate, and thus the pump transition is strongly saturated,
 the resulting Lorentzian widths of the DR transitions are found to follow simple expressions, which greatly
 simplifies the physical interpretation of such experiments.  The width of the DR features are different for the
 pump and probe field co- and counter-propagating\cite{Feld69} and this can be used to distinguish the two DR
 transitions in the case of a standing wave probe field.

.
 
 \section{Theory of Double Resonance Lineshapes}
\begin{figure}[h]
\begin{center}
\includegraphics[width=10cm]{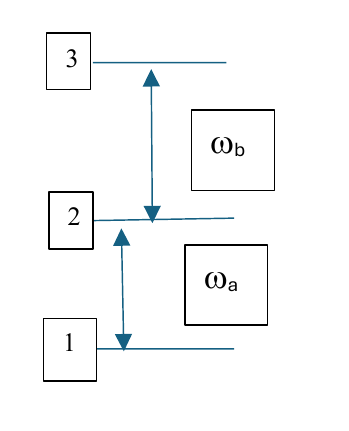}  
\caption{Level Diagram for Ladder-type Double Resonance}
\label{fig:level_diagram}
\end{center}
\end{figure}

 Consider the energy diagram  for double resonance (DR) shown in figure~\ref{fig:level_diagram} with three states $1-3$, with the energy of state $n$ given by $E(n) = \hbar\omega_n$. 
 We will first consider the case of ladder-type DR (LDR) where $E(1) < E(2) < E(3)$ and the states are connected by 1-photon allowed transitions  $1 \leftrightarrow 2 $ and $2 \leftrightarrow 3$, and the thermal population is assumed to be entirely in state 1.  This case has also
 been called cascade-type in the literature.~\cite{Berman11}
 Later, we will discuss changes in the results for treatment of V-type DR where $E(2) < E(1), E(3)$ and the thermal population is all in state $2$, and  also $\Lambda$-type DR where $E(1) < E(3) < E(2)$.  We will call the radiation nearly resonant with the $1 \leftrightarrow 2$ transition the pump and indicate its angular frequency by $\omega_a$, its wavenumber by $k_a$,
 and its field amplitude by ${\cal \vec E}_a$ and (with $a$ replaced by $b$) for the corresponding values for the probe  wave that is nearly resonant with the $2 \leftrightarrow 3$ transition. 
 
Herein, we focus on the experimental case where the pump radiation significantly saturations its transition while the probe field is well below saturation, but the results presented will only be in that limit when explicitly stated.  We make the usual rotating-wave approximation and define the two detuning parameters $\Delta \omega_{12} = \omega_2-\omega_1 - \omega_a$ and $\Delta \omega_{23} = \omega_3 - \omega_2 - \omega_b$ and Rabi frequencies 
$\Omega_{12} =  \, \left<2| \vec \mu |1 \right>\cdot{\cal \vec E}_{a} / \hbar$ and $\Omega_{23} =  \, \left<3| \vec \mu |2 \right>\cdot{\cal \vec E}_{b} / \hbar$ for the pump and probe transitions.  $\vec \mu$ is the electric dipole moment operator for the absorber.
Further, we assume that the field at $\omega_a$ only drives the $1 \leftrightarrow 2$ transition and that at $\omega_b$ only drives the $2 \leftrightarrow 3$ transition and we neglect 
nonresonant light shifts.

 For DR experiments with monochromatic pump waves, only a small part of the $v_z$ distribution of state 1 is pumped leading to a Bennett hole in the distribution of state 1 and
 Bennett hill in state 2.\cite{Bennett62}   When $\Omega_{12}$ is much less than the Doppler width of its transition, as is the case for the experiments we seek to model, even low angle scattering will detune coherence away from resonance and dephase the absorption.
 As a result, much like Lamb-dips,  the relaxation will be dominated by the
total collision rate, including elastic collisions, which is determined by the long rang attractive interaction between the collision partners.\cite{Child74}  As an example, Barger and Hall\cite{Barger69} measured the pressure broadening of the Lamb dip of the CH$_4$ $\nu_3$ P(7) line that overlaps with the 3.39 $\mu$m He-Ne laser and found the self-broadening rate to be 16.4 MHz/torr, which implies a relaxation rate $2\pi$ times larger, or 1.03E8/torr.   Using the expression given by Child\cite{Child74} for the collisional cross section versus collisional velocity and neglecting the correction that arises from Glory scattering, integration over collisional velocity, one finds the collisional rate to be
\begin{equation}
{\rm Collision \,\, rate} = N \int \sigma(v) v P_{MB}(v) dv = N \times 10.481 \left( \frac{C_6}{\hbar}    \right)^{2/5} \left(  \frac{k_{\rm B}T}{\mu}  \right)^{3/10}
\end{equation}
where $N$ is the number density of the gas, $\sigma(v)$ the collision cross section as function of center of mass collision velocity, $C_6$ the coefficient of the long range interaction ( = $C_6/r^6$),
$T$ the sample gas temperature, and $\mu$ collision reduced mass.  The neglected Glory contribution is oscillatory with collision speed and largely washes out after the thermal average over collsion speed.   Using the methane-methane long range coefficient $C_6$ = 129.6 $E_{\rm H} a_0^6$.\cite{Thomas77}, and $T = 300$\,K the total collision rate of 1.00E8/torr, within 3\% of the Lamb Dip broadening calculated rate.  The Lamb-dip derived collision rate corresponds to a hard-sphere collision radius of 1.07 nm at which the Lennard-Jones potential is only 0.84\% of the well depth (148.2 $k_{\rm B}$).  In contrast,  HITRAN\cite{Gordon2026} lists the self-broadening rate of this CH$_4$ line as 3.08 MHz/torr, only 19\% of the Lamb-Dip rate and comparable a hard-sphere collision rate ($\sigma = 0.382$\,,nm)/2$\pi$ of 2.1 MHz/torr.   The pressure broadening coefficients commonly found in the literature and databases such as HITRAN are dominated by inelastic collisions and correspond to collisions rates far less than the total collision rate including elastic rates.\cite{Rabitz74}   In traditional pressure broadening theory, such as Anderson theory and its extensions, elastic collisions lead to pressure shift coefficients and pure dephasing, but instead lead to population relaxation when molecular velocity is included among the quantum numbers as required for treatment of velocity selected excitation.  

 We will make the assumption that all populations and coherences have equal relaxation rate, $\gamma$, which simplifies the expressions considerably.  The relaxation rate is proportional the the total collision rate, and the later scales of $C_6^{2/5}$ power.   The $C_6$ coefficient is proportional to the product of the electric polarizabilities of the collision partners.  The electric polarizability typically changes by less than 1\% per vibrational quantum.  See, for example, values for H$_2$O, NH$_3$, and CH$_4$ of the zero-point contribution to the electric polarizability of different modes given
 by O. Quinet, B. Champagne, and B. Kirtman\cite{Quinet01}.  Thus we anticipate that assumption of equal relaxation rates is an excellent approximation for Doppler selected Double Resonance vibrational spectroscopy.

We will consider the constant field amplitude, steady-state solutions for the density matrix, $\rho$, which should be appropriate if the mean free path of absorbers is much less than the beam radii of the optical fields.   We can always select the phases of the states such that $\Omega$'s are real and positive.   In the general case, one would need to consider the  distribution of Rabi frequencies for the different $M_J$ states of a given initial level, but that will result in a lineshape that is the appropriate average over those for the individual initial $M_J$ values,~\cite{Lehmann23} so this complication will be neglected in this work.    We will initially neglect Doppler broadening of the transitions, and then later consider the result of convoluting the resulting excitation lineshape over Doppler broadening.
  
  The steady-state probe field photon absorption rate per molecule thermally in state 1 will equal to the rate that molecules in state 3 decay times the fractional population in state 3 and thus will be proportional to $\gamma \rho_{33}$.   The steady-state pump absorption will be proportional to $\gamma( \rho_{22} + \rho_{33})$.  DR can also be observed by monitoring spontaneous emission from states 2 or 3, which have rates proportional to the spontaneous emission rates times the corresponding density matrix population term.   Explicit time-evolution expressions (Optical Bloch equations) for the 3-level density matrix can be found in many publications.\cite{Hansch70, Feneuille75, Salomaa75, Berman11}   However, many of these include explicit population transfer terms between levels connected by spontaneous emission and may even neglect
  collisional relaxation terms.   So, for clarity, we give the steady-state equations for the above assumptions in the field-interaction representation which eliminates all the explicit time dependence from the effective Hamiltonian:
  \begin{equation}
H = 
\hbar \left(
\begin{array}{ccc}
 -\Delta\omega_{12} & -\Omega_{12}/2  &  0 \\
-\Omega_{12}/2    & 0  &   -\Omega_{23}/2 \\
 0  &  -\Omega_{23}/2  &    \Delta\omega_{23} 
\end{array}
\right)
\end{equation}
The Liouville equation for the time dependence of the density matrix is:
\begin{equation}
\frac{d\rho}{dt} = \frac{1}{i\hbar} \left[ H, \rho \right] - \gamma (\rho - \rho^{\rm e} )
\end{equation}
where $\rho^{\rm e}$ is the equilibrium density matrix for the three level system.  We will take $\rho_{ij}^{\rm e} = \delta_{i,1}\delta_{j,1}$ for ladder-and $\Lambda$-type DR and $= \delta_{i,2}\delta_{j,2}$
for V-type DR.
The steady-state conditions $d\rho/dt = 0$ lead to equations 
 \begin{eqnarray}
  \gamma( \rho_{11} - \rho_{11}^e)  + i (\Omega_{12}/2) (\rho_{12} - \rho_{12}) &=& 0 \nonumber \\
 i (\Omega_{12}/2) (\rho_{11}-\rho_{22}) + (\gamma - i \Delta\omega_{12}) \rho_{12} + i (\Omega_{23}/2) \rho_{13} &=& 0 \nonumber \\
 i (\Omega_{23}/2) \rho_{12} + (\gamma - i (\Delta\omega_{12} + \Delta\omega_{23})) \rho_{13} - i (\Omega_{12}/2) \rho_{23} &=& 0 \nonumber \\
 -i (\Omega_{12}/2) (\rho_{11}-\rho_{22}) + (\gamma+ i \Delta\omega_{12}) \rho_{12}  - i (\Omega_{23}/2) \rho_{31}  &=& 0 \nonumber \\
 -i (\Omega_{12}/2) (\rho_{12} - \rho_{12}) + \gamma (\rho_{22}-\rho_{22}^e) + i (\Omega_{23}/2) (\rho_{23} - \rho_{32})  &=& 0 \nonumber \\
 -i (\Omega_{12}/2) \rho_{13} + i(\Omega_{23}/2) (\rho_{22}-\rho_{33}) + (\gamma - i \Delta\omega_{23}) \rho_{23}  &=& 0 \nonumber \\
  -i(\Omega_{23}/2) \rho_{12} + (\gamma + i (\Delta\omega_{12} + \Delta\omega_{23}) ) \rho_{31} + i (\Omega_{12}/2) \rho_{32}  &=& 0 \nonumber \\
  -i(\Omega_{23}/2) (\rho_{22}-\rho_{33}) + i (\Omega_{12}/2) \rho_{31} + (\gamma + i \Delta\omega_{23}) \rho_{32}  &=& 0 \nonumber \\
  -i(\Omega_{23}/2) (\rho_{23} - \rho_{32}) + \gamma (\rho_{33} -\rho_{33}^e)  &=& 0   \label{eq:rho_ss_eq}
\end{eqnarray}

 Using Mathematica, analytical expressions for the  steady state solutions to Eqs.~\ref{eq:rho_ss_eq} were found but are rather complex and opaque and are listed in the
 Appendix.  These solutions are considerably simplified if one makes the weak-probe field limit where we neglect all but the lowest power in $\Omega_{23}^2$.  For the case of
 Ladder type DR
   \begin{eqnarray}
 \rho_{22}^{\rm wp}&=&  \frac{\rm NUM_{22}^{\rm wp}}{\rm DEN^{\rm wp}} \hspace{0.5in} \rho_{33}^{\rm wp} =  \frac{\rm NUM_{33}^{\rm wp}}{\rm DEN^{\rm wp}}  \hspace{0.5in}   \nonumber \\  
{\rm NUM_{22}^{wp}} &=& \Omega_{12}^2 \left( 16 \gamma^4 + \left( -4 \Delta\omega_{23} (\Delta\omega_{12}+\Delta\omega_{23}) + \Omega_{12}^2 \right)^2 
   +  8 \gamma^2 \left(  2 \Delta\omega_{12}^2 + 4\Delta\omega_{12}\Delta\omega_{23} + 4 \Delta\omega_{23}^2  + \Omega_{12}^2\right) \right. \nonumber \\
 && \left.   + 2 \left( 2(\Delta\omega_{12}+\Delta\omega_{23}  )(3\Delta\omega_{12}+4\Delta\omega_{23})+ \Omega_{12}^2 + 4 \gamma^2 \right) \Omega_{23}^2  \right) \nonumber \\
{\rm NUM_{33}^{wp}}  &=& \Omega_{12}^2 \Omega_{23}^2 \left(12 \gamma^2
   +4 \left(\Delta \omega_{12}^2
   +\Delta \omega_{12} \Delta \omega_{23}+
   \Delta \omega_{23}^2\right)
   +3 \Omega_{12}^2 \right) \nonumber \\
   {\rm DEN^{wp}} &=& 
   2 ( \gamma^2+ \Delta \omega_{12}^2 + \Omega_{12}^2) \nonumber \\
  & & \left( 16 \gamma^4 + 16 (\gamma^2 +  \Delta \omega_{23}^2) (\Delta \omega_{12}+\Delta \omega_{23})^2
   +16 \gamma^2 \Delta\omega_{23}^2  \right. \nonumber  \\
  && \left. +8 \Omega_{12}^2 (\gamma^2 - \Delta \omega_{23}(\Delta \omega_{12} + \Delta \omega_{23}) ) + \Omega_{12}^4 \right)  \label{eq:DENwp}
  \end{eqnarray}  
   The superscript, wp, indicates it is the weak probe field limit.  The expressions are further simplified if we take the lowest order terms in both $\Omega_{12}, \Omega_{23}$
  \begin{eqnarray}
\rho_{22}^{\rm wpp} &=& \frac{\Omega_{12}^2}{2\left(\gamma^2 + \Delta\omega_{12}^2 \right)} \nonumber \\
&&
- \frac{ \left( 4 \gamma^4 + \Delta\omega_{12} (  \Delta\omega_{12}+  \Delta\omega_{23}) (  \Delta\omega_{12}^2-2 \Delta\omega_{23}^2 )
+ \gamma^2 (  \Delta\omega_{12}^2 -  \Delta\omega_{12}  \Delta\omega_{23} + 2  \Delta\omega_{23}  ) \right) \Omega_{12}^2 \Omega_{23}^2 }
{8(\gamma^2+\Delta\omega_{12}^2  )^2 (\gamma^2 + \Delta\omega_{23}^2) (\gamma^2 + (\Delta\omega_{12}+\Delta\omega_{23})^2 ) } \nonumber \\
\rho_{33}^{\rm wpp} &=& \frac{ \left( 3 \gamma^2 + \Delta\omega_{12}^2 + \Delta\omega_{12} \Delta\omega_{23} + \Delta\omega_{23}^2\right) \Omega_{12}^2 \Omega_{23}^2  }
{8(\gamma^2+\Delta\omega_{12}^2  ) (\gamma^2 + \Delta\omega_{23}^2) (\gamma^2 + (\Delta\omega_{12}+\Delta\omega_{23})^2 ) }
\end{eqnarray}
 Here, the superscript wpp stands for weak pump and probe.  There are resonance peaks when $\Delta\omega_{12}, \Delta\omega_{23},$ and $ \Delta\omega_{12}+\Delta\omega_{23} = 0$,
 corresponding to pump, probe and two-photon resonances.  At exact double resonance, $\Delta\omega_{12} = \Delta\omega_{23} = 0$, $\rho_{22}^{\rm wpp} \rightarrow 
 \frac{\Omega_{12}^2}{2 \gamma^2} \left( 1 - (\Omega_{23}/\gamma)^2 \right)$ and $\rho_{33}^{\rm wpp} \rightarrow 3 \Omega_{12}^2 \Omega_{23}^2/8\gamma^4$.

\begin{figure}[h]
\begin{center}
\includegraphics[width=10cm]{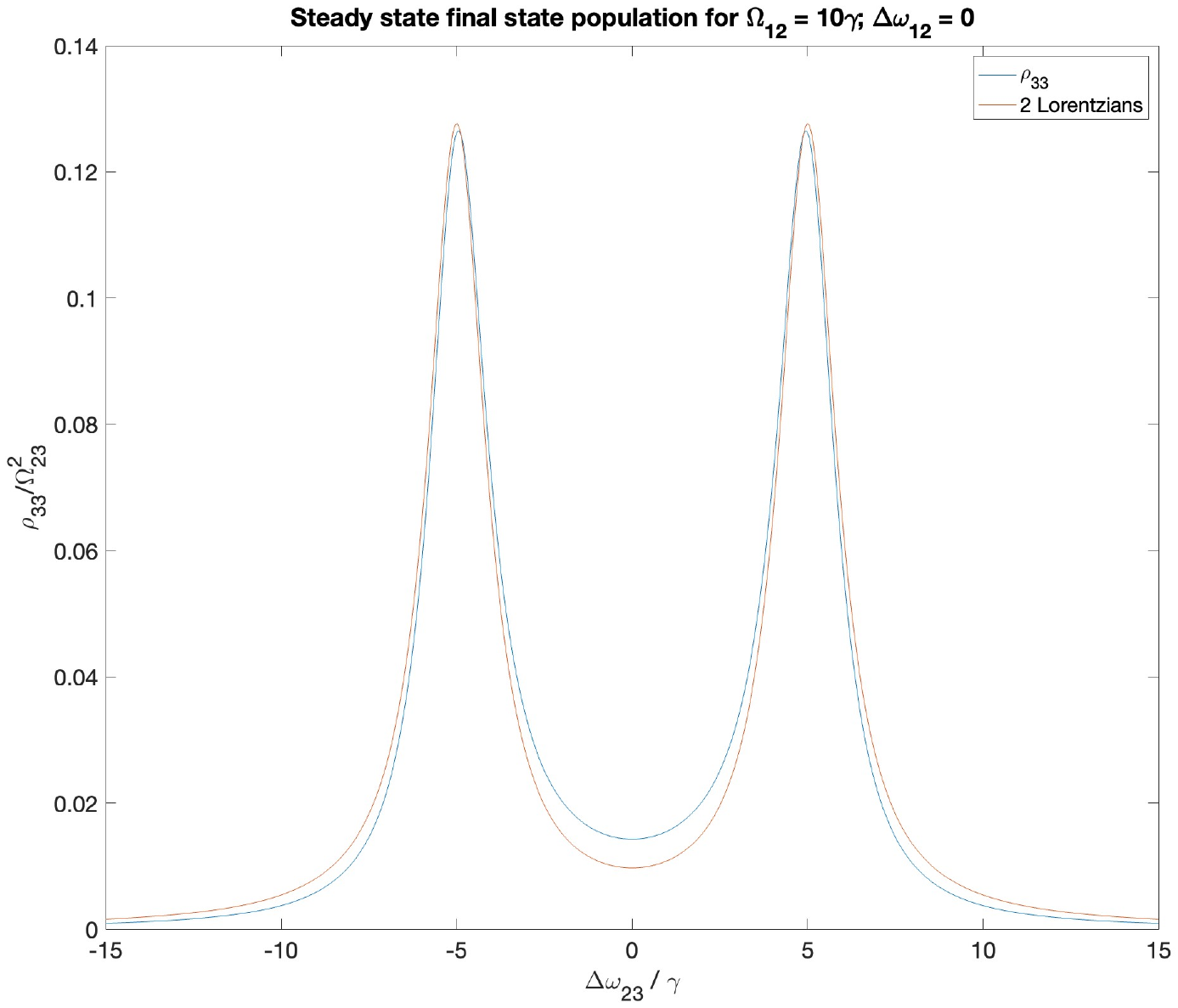}  
\caption{Ladder-type, weak probe Double Resonance Excitation Spectrum, $\rho_{33}^{\rm wp}/\Omega_{23}^2$,  for on-resonance pumping, $\Delta \omega_{12} = 0$, and zero Doppler width.}
\label{fig:On-Resonane_pumping}
\end{center}
\end{figure}

Figure~\ref{fig:On-Resonane_pumping} displays the excitation rate in the wp limit as a function of probe detuning, $\Delta \omega_{23}$, with $\Omega_{12} = 10 \gamma$ and $\Delta \omega_{12} = 0$.   The spectrum consists of a pair of peaks centered at $\Delta \omega_{23} = \pm \Omega_{12}/2$.  Also displayed is the sum of two Lorentzians centered at these positions with half-width at half maximums equal to $\gamma$.   This result has long been known\cite{Feneuille75} and is easily rationalized as due to the Autler-Townes effect, a.k.a. AC Stark Effect.\cite{Autler55}  The Rabi flopping of the pump transition leads to states 1 and 2 producing a pair of dressed time-independent mixed states split by $\Omega_{12}$.\cite{CohenTannoudji69, Berman11}    The sums of Lorentzians are a good approximation, but they slightly underestimate the absorption near $\Delta \omega_{23} \approx 0$; this is a result of an interferences of the transition amplitudes of the two dressed states.   We note that the probe transition is NOT power broadened by the pump, each dressed state peak has the expected homogeneous width, $\gamma$.

\begin{figure}[h]
\begin{center}
\includegraphics[width=10cm]{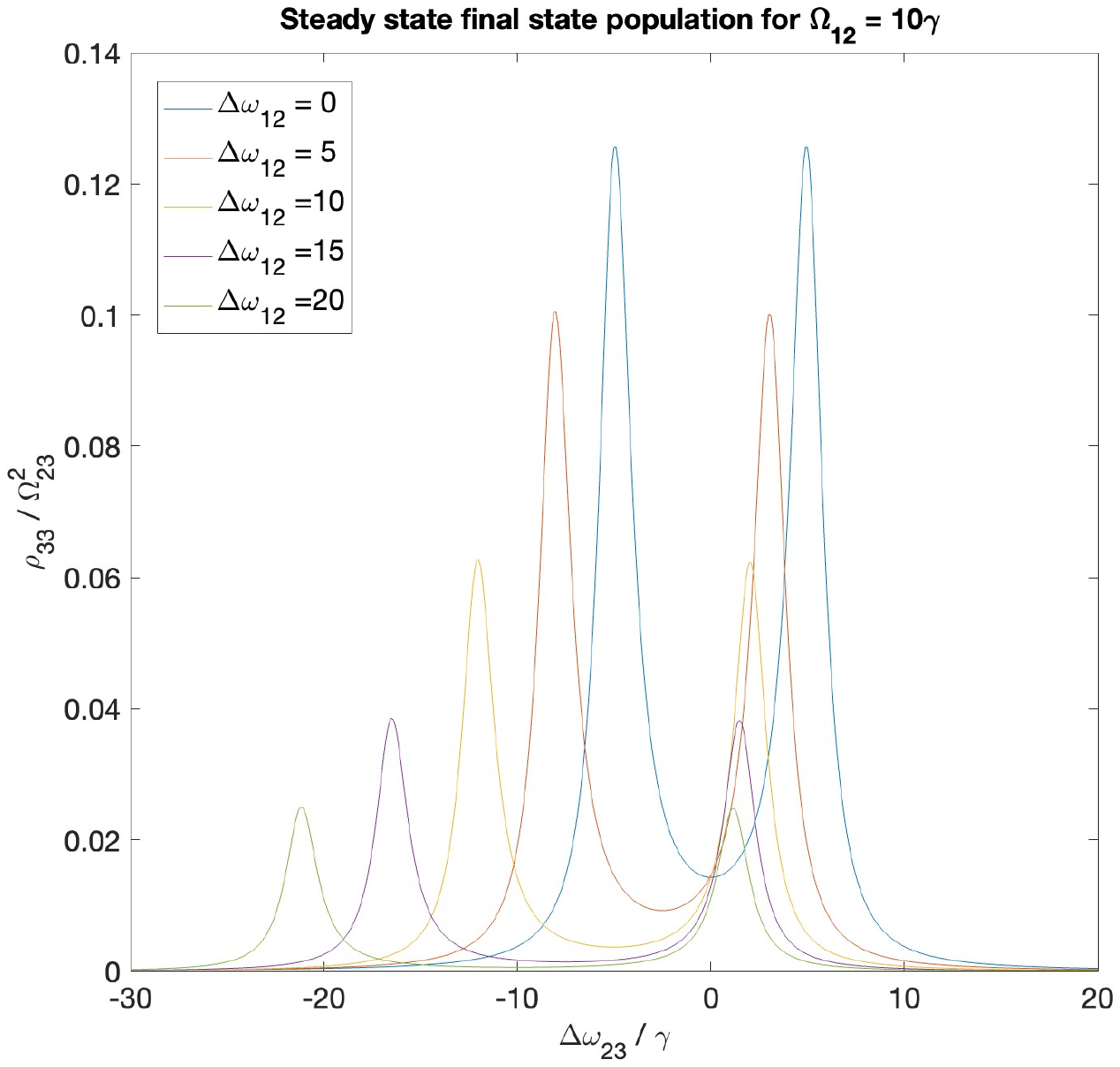}  
\caption{Ladder-type, weak probe Double Resonance Excitation Spectrum, $\rho_{33}^{\rm wp}/\Omega_{23}^2$,  for multiple pump detuning values,  $\Delta \omega_{12} $, and zero Doppler width.}
\label{fig;multiple_pumping_detuning}
\end{center}
\end{figure}

Figure~\ref{fig;multiple_pumping_detuning} displays the excitation spectrum under the same conditions but with multiple values of $\Delta \omega_{12}$.  For each case, the excitation spectrum is again the sum of two equal height peaks with nearly Lorentzian lineshapes and half-width at half maximum, HWHM, of $\gamma$.  The centers of the peaks are at $\Delta \omega_{23} =  - (\Delta \omega_{12} \pm \sqrt{ \Delta \omega_{12}^2 + \Omega_{12}^2} )/2$.  This is exactly the shift expected from the positions of the Dressed states.\cite{Berman11}   For $\Delta \omega_{12}^2 >> \Omega_{12}^2$, one peak moves to $\Delta \omega_{23} = 0$, {\it i.e.}~the unperturbed probe resonance frequency and the other to $-\Delta \omega_{12}$, the expected position of the $1 \leftrightarrow 3$ two-photon resonance.  For $\Delta \omega_{12} < 0$, the excitation spectrum as a function probe detuning is just the mirror image of that with positive pump detuning.  The one-photon resonance near zero probe detuning arise from state $2$ being pumped in the wing of its absorption line.  

We can define the two dressed states\cite{Berman11} as $\left|\pm\right>$ which can be written in terms of states $1$ and $2$  by $\left|+\right>= \cos(\theta)\left|1\right> + \sin(\theta\left|2\right>$ and $\left|-\right> = -\sin(\theta)\left|1\right> + \cos(\theta)\left|2\right>$ with $\tan(2\theta) = \Omega_{12}/\Delta\omega_{12}$.  The thermal population in the
dressed states are proportional to $|\left<\pm|1\right>|^2$ and the absorption to state 3 per molecule in each dressed state proportional to $|\left<\pm|2\right>|^2$, so in both cases we have probe absorption strength proportional to $\left<\pm|1\right>^2 \left<\pm|2\right>^2 = \sin(\theta)^2 \cos(\theta)^2 = \sin(2\theta)^2/4
= \Omega_{12}^2/4(\Delta \omega_{12}^2 + \Omega_{12}^2)$.

\begin{figure}[h]
\begin{center}
\includegraphics[width=10cm]{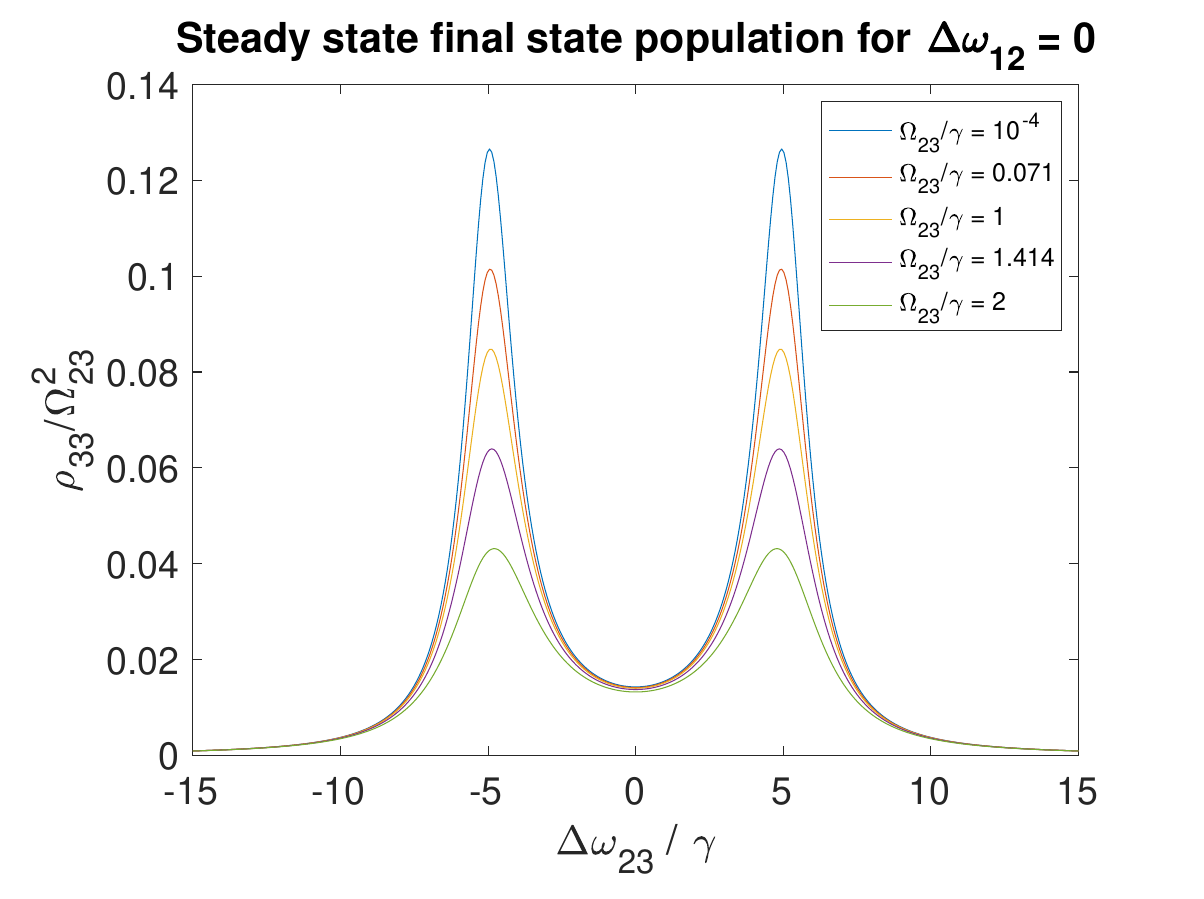}  
\caption{Ladder-type Double Resonance Excitation Spectra, $\rho_{33}/\Omega_{23}^2$ for on resonance pumping, $\Delta \omega_{12} = 0$, pump level $\Omega_{12} = 10\gamma$, with zero Doppler width, and at multiple probe power levels.}
\label{fig;Excitation_on_resonance_multiple_R23}
\end{center}
\end{figure}

\begin{figure}[h]
\begin{center}
\includegraphics[width=10cm]{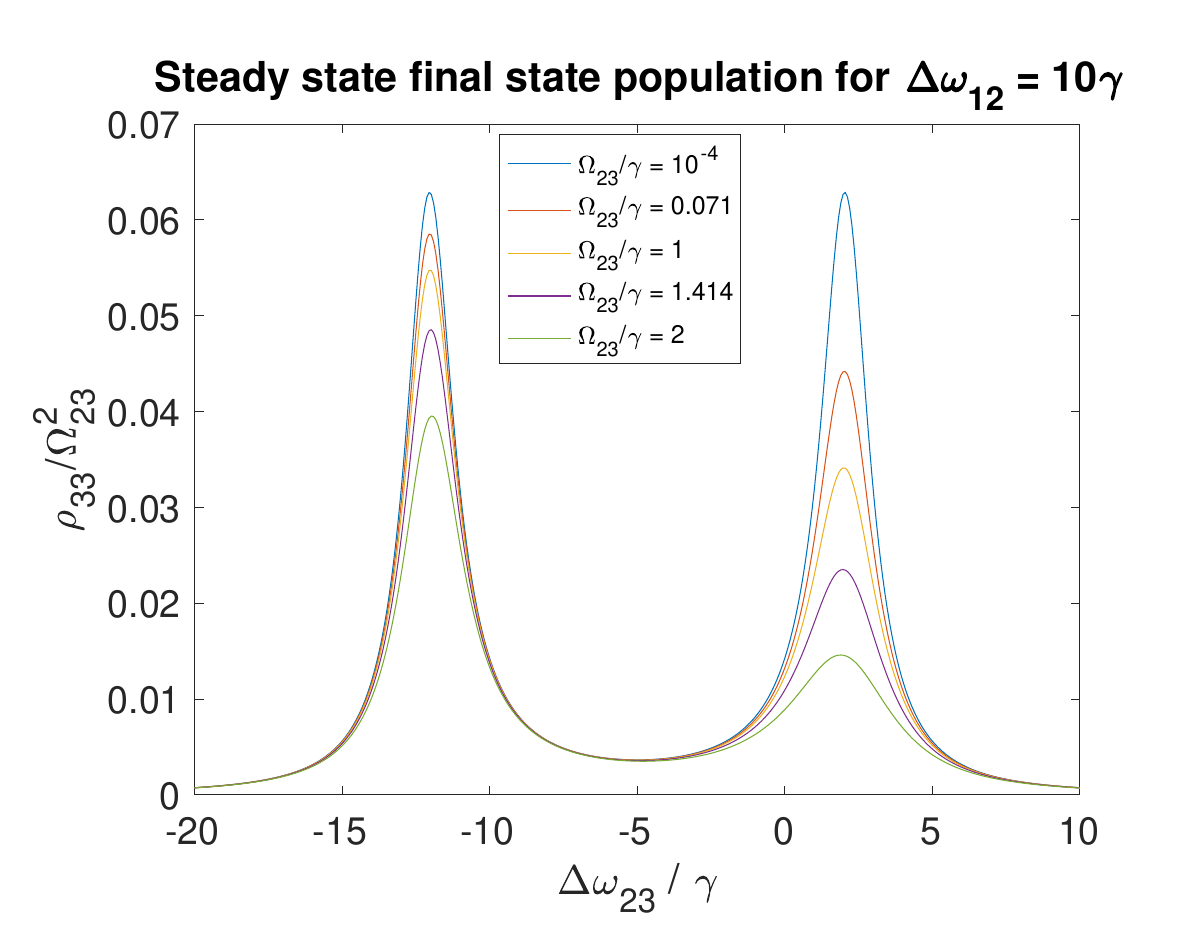}  
\caption{Ladder-type Double Resonance off-pump resonance Excitation Spectra, $\rho_{33}/\Omega_{23}^2$, for off-resonance pumping,  $\Delta \omega_{12} = 10 \gamma$, $\Omega_{12} = 10 \gamma$, with zero Doppler width, and at multiple Probe power levels. }
\label{fig;Excitation_off_resonance_multiple_R23}
\end{center}
\end{figure}

Figure~\ref{fig;Excitation_on_resonance_multiple_R23} shows the calculated excitation spectra for several values of $\Omega_{23}$ without making the weak probe approximation.  For $\Delta \omega_{12} = 0$, both peaks saturate at the same rate with a saturation intensity about twice that expected for the unperturbed probe transition; the saturation index\cite{DemtroderLS2}  $ S = 1 $ for $\Omega_{23}^2 \approx 2\gamma^2$.  This is also easily rationalized as each dressed state has 50\% state 2 character when the pump is on-resonance.    For a detuned pump field, the peak moving towards $\Delta \omega_{23} = 0$ saturates more easily while the other saturates more slowly.  For $\Delta \omega_{12}^2 >> \Omega_{12}^2$, $ S = 1 $  for $\Omega_{23}^2 \approx \gamma^2$, the $\Delta \omega_{23} \rightarrow 0$ peak has the same saturation parameter as the unperturbed probe transition.   This is as expected as the lower state of this transition is the dressed state correlated with pure state $2$ for large pump detuning. The two-photon resonance requires increasing intensity to saturate as detuning increases, with $S$ inversely
proportional to $\Delta\omega_{12}^2$.  This asymmetry in saturation is demonstrated in Figure~\ref{fig;Excitation_off_resonance_multiple_R23}

For $\Lambda$-type DR, the same expressions hold with the change that we flip the sign of $\Delta \omega_{23}$. as now $\Delta \omega_{12}$ and $\Delta \omega_{23}$ need to tune in the same direction to stay in two-photon resonance.  Also, in this case, transition to state $3$ leads to gain on the probe transition.

\clearpage

\section{DR Lineshape with Doppler Broadened Transitions}
We now introduce Doppler broadening.  For ladder-type DR and for each value of $v_z$ ($z$ axis is parallel to the pump propagation direction), we must replace $\Delta \omega_{12}$ by 
$\Delta \omega_{12} + k_a v_z$ and $\Delta \omega_{23}$ by $\Delta \omega_{23} \pm k_b v_z$, in the above expressions for $\rho_{ij}$,  with the $+$ sign for the probe field co-propagating with the pump, and the negative sign if it counter-propagates.   For each $v_z$, we get a spectrum dominated by two Lorentzian lines but to calculate the thermal probe spectrum, we need to
average the resulting lineshape over the thermal Gaussian distribution of $v_z$ values which has standard deviation $u = \sqrt{k_BT/M}$.\cite{Berman11}   We will take the limit where the Doppler width, $k_a u$, is much greater than $ \gamma, \Omega_{12}$
which allows us to replace
\begin{eqnarray}
&& \int_{-\infty}^{\infty} g_D(\Delta \omega_{12} + k_a v_z) \rho_{33}(\Delta \omega_{12} + k_a v_z, \Delta \omega_{23} \pm k_b v_z )  dv_z \label{eq:Doppler_int}\\
&&\approx g_D(\Delta \omega_{12}) \int_{-\infty}^{\infty}  \rho_{33} (\Delta \omega_{12} + k_a v_z, \Delta \omega_{23} \pm k_b v_z )  dv_z   \label{eq:infiniteD}
\end{eqnarray}
which removes the Doppler broadening  from effecting the DR lineshape except for a scale factor.  This is a good approximation for room-temperature ro-vibrational spectra even with pump intensities up to $100$\,W/cm$^2$.

Even with Mathematica's help, we were not able to evaluate analytically the resulting integral of Eq.~\ref{eq:infiniteD} using $\rho_{33}$ or even $\rho_{33}^{\rm wp}$ but were able to
do so using $\rho_{33}^{\rm wpp}$.  In that case, the result was Lorentzian lineshapes in $\Delta\omega_{23}$ centered on $\Delta\omega_{23} = \pm \Delta\omega_{12} (k_b/k_a)$
with + being the co- and - the counter-propagating probe.   For co-propagating, the HWHM = $(1+ k_b/k_a) \gamma$; for counter-propagation, the HWHM = $\gamma$ for $k_b < k_a$
and = $(k_b/k_a) \gamma$ for $k_b > k_a$.   Co- and Counter-propagating peaks have equal area, which is independent of $k_b/k_a$.

For the case of a saturating pump, of more interest with regards to experiments, we have numerically integrated Eq.~\ref{eq:infiniteD} for a range of cases which
are defined by $\Omega_{12}/\gamma$, $\Omega_{23}/\gamma$ and $k_b/k_a$.   The numerical results
reproduce the analytical result in the weak pump and probe (wpp) limit of $\Omega_{12}, \Omega_{23} << \gamma$.   
In the strong pumping but weak probe power limit (wp above), $\Omega_{12} >> \gamma >> \Omega_{23}$ and for $k_b > k_a$ the DR lineshapes  remain Lorentzian with amplitude proportional to  $\Omega_{23}/\gamma$  and  HWHM  = $(k_b / k_a + 0.5) \gamma$
and $(k_b / k_a - 0.5) \gamma$ for co- and counter-propagating waves.  Again, the co- and counter- peaks have equal areas.  The asymmetries of the peak widths for the two probe propagation directions was noted by Feld and Javan.\cite{Feld69}

\begin{figure}[h]
\begin{center}
\includegraphics[width=17cm]{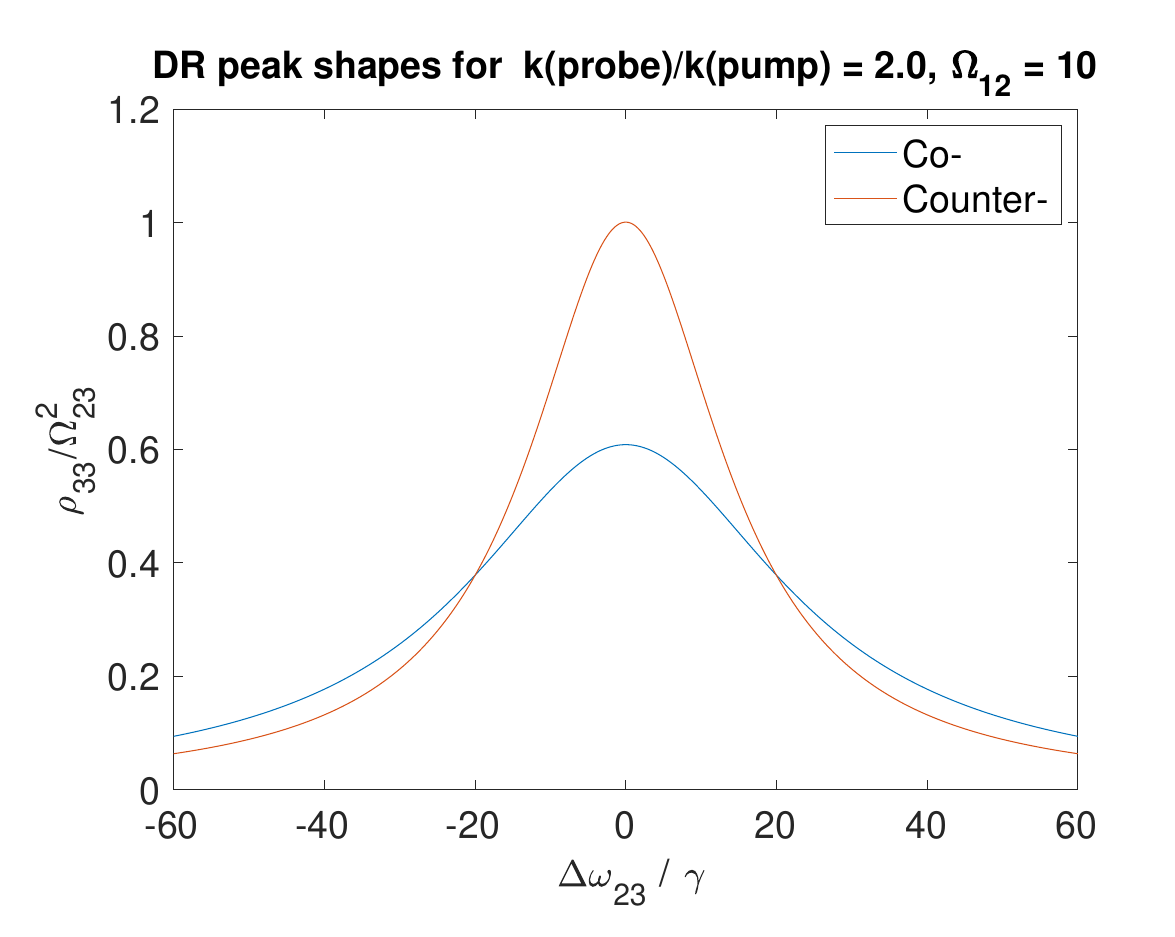}  
\caption{Calculated Ladder-type DR lineshape after convolution over Doppler detuning in the weak probe limit, $\Omega_{23} << \gamma$, calculated with $\Omega_{12} = 10\gamma$, $\omega_b = 2 \omega_a$, and $\Delta \omega_{12} = 0$.  For pump detuned from center of the line, the co-propagating probe peak will be centered as $\Delta \omega_{23} = (\omega_b/\omega_a) \Delta \omega_{12}$ while the counter-propagating  peak will be centered at $\Delta \omega_{23} = -(\omega_b/\omega_a) \Delta \omega_{12}$.}
\label{fig:DB_DR_R12=10_k2=2k1}
\end{center}
\end{figure}

We present graphic results for the case where $\Delta \omega_{12} = 0$ and  $k_b = 2 k_a$ as this corresponds to the  CH$_4$ ground vibrational state $\leftrightarrow \nu_3 \leftrightarrow 3\nu_3$ DR experiments we have been
 conducting that motivated the present investigation.
Figure~\ref{fig:DB_DR_R12=10_k2=2k1} shows the weak-probe field co- and counter-propagating DR peaks for $\Omega_{12} = 10 \gamma$.   Both spectra appear to be single Lorentzian peaks with HWHM approximately 2.5 and 1.5 $\Omega_{13}$ for the co- and counter-propagating cases.   The extra width for the co- verses counter-propagation probe is easily rationalized when one recalls that the Doppler shift of a two-photon resonance is proportional to $k_b + k_a$ in the co- and $|k_b - k_a|$ in the counter-propagating cases.   These widths flip for the $\Lambda$-type DR as do the two-photon Doppler widths as then the probe transition is stimulated emission instead of absorption.  When making the infinite Doppler width approximation, Eq.~\ref{eq:infiniteD}, the residuals of a fit to the DR transition to a Lorentzian are limited by the convergence of the numerical integral, maximum values less than $10^{-10}$ for this case.  If one does not make the infinite Doppler width approximation, the calculated spectrum has wings that fall off faster than the best fit Lorentzian as this part of the probe spectrum (as well as the region of $\Delta \omega_{23} \approx 0$) arise from absorbers with large Doppler detuning and thus are suppressed by the Doppler lineshape.  For the detuned pump radiation case, the centers of the DR peaks shift by $\pm (k_b/k_a) \Delta \omega_{12}$ for the co- and counter-propagating cases for ladder-type DR; the opposite for $\Lambda$-type.  The width and height of the DR peaks do not change with $\Delta\omega_{12}$ as long as the detuning is sufficiently small that the approximation in Eq.~\ref{eq:infiniteD} can be made, but they are centered at $\mp (k_b/k_a) \Delta \omega_{12}$ for co- and counter-propagating probe wave.

 \clearpage
 
\begin{figure}[h]
\begin{center}
\includegraphics[width=17cm]{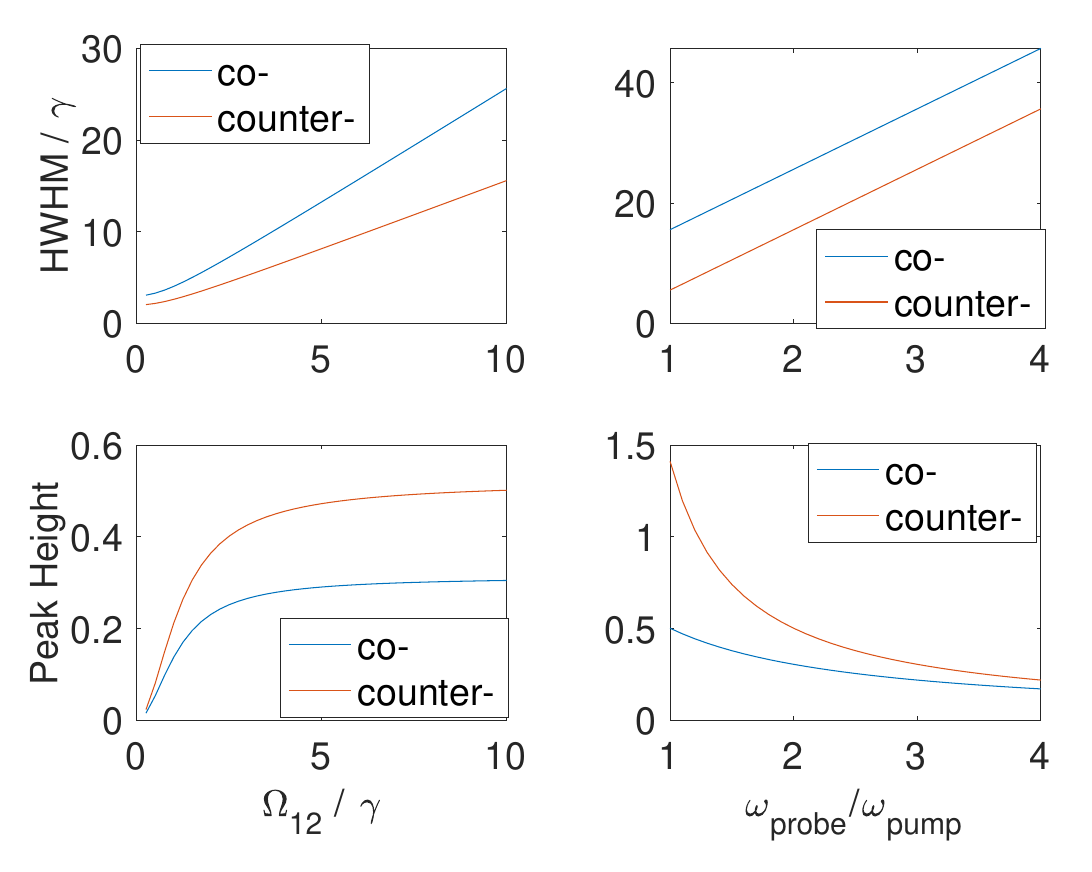}  
\caption{Lorentzian weak probe ladder-type DR peak HWHM (in units of relaxation rate $\gamma$) and peak heights as a function of Rabi frequency of the pump transition, $\Omega_{12}$, for $\omega_b = 2 \omega_a$ and
as a function of $\omega_b / \omega_a$ for $\Omega_{12} = 10 \gamma$.  The Doppler width is assumed  $>> \Omega_{12}$}
\label{fig:L-type_Doppler_peaks}
\end{center}
\end{figure}

 We calculated the weak probe Doppler convoluted spectra for a range of pump powers, $\Omega_{12} = (0.1 - 10)\gamma$ for the case that $k_b = 2 k_a$ and also for $k_b = 1-4 \times k_a$ with $\Omega_{12} = 10 \gamma$.  Each Ladder-type DR peak was fit to a Lorentzian lineshape.  The Lorentzian peak HWHM and peak heights are displayed in Figure~\ref{fig:L-type_Doppler_peaks}.
 These plots demonstrate the linear relationship between width and both pump Rabi frequency and $k_b/k_a$ in the strong pumping limit. 
 The linear dependence of the width on pump Rabi frequency is  qualitatively similar to the broadening with increasing pump power of a Lamb dip or Bennett hole and such
 transitions are often described as power broadened.\cite{Salomaa75, Schuurmans77}    
 We note that the square of the DR peak HWHM is not linear in $\Omega_{12}^2$ as it is for Lamb dip.\cite{DemtroderLS2}   
 For $\Omega_{12} = 10$ and $k_b = 2 k_a$ the HWHM's are 25.625 and 15.575$\gamma$, while the standard scaling of the Lamb dip would predict widths of 25.080 and 15.033 $\gamma$ for co- and counter-propagating cases; the widths converge more slowly to the high power limits which predict 25 and 15$\gamma$ respectively.

\begin{figure}[h]
\begin{center}
\includegraphics[width=10cm]{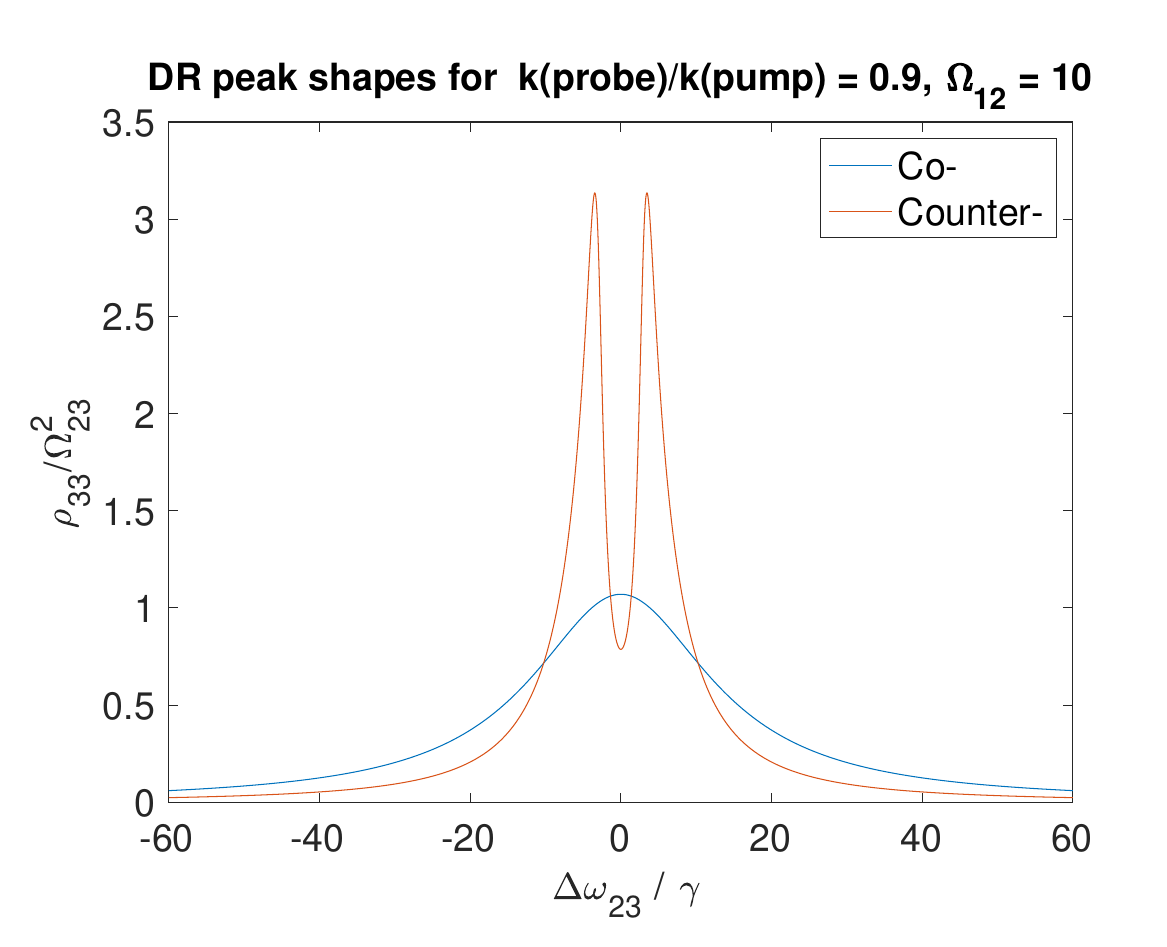}  
\caption{Doppler averaged weak probe Ladder-type Double Resonance for $\Delta\omega_{12} = 0,  \Omega_{12} = 10 \gamma$ and $ \omega_b = 0.9 \omega_a$.  The signal from the co-propagating probe wave remains Lorentzian but the counter-propagating probe wave retains a Autler-Townes type of splitting after averaging.}
\label{fig:DR_R12=10_k2=0.9k1}
\end{center}
\end{figure}

  For  $k_b < k_a$, the co-propagating DR peak remains a single Lorentzian but the counter-propagating peak splits into two peaks and is strongly deviated from a Lorentzian lineshape.   
  Figure~\ref{fig:DR_R12=10_k2=0.9k1} displays the calculated co- and counter-propagating DR spectra for $\Omega_{12} = 10 \gamma$ and $k_b = 0.9 k_a$.  The
  strong deviation of the Counter-propagation probe spectrum from Lorentzian lineshape is evident.

\begin{figure}[h]
\begin{center}
\includegraphics[width=10cm]{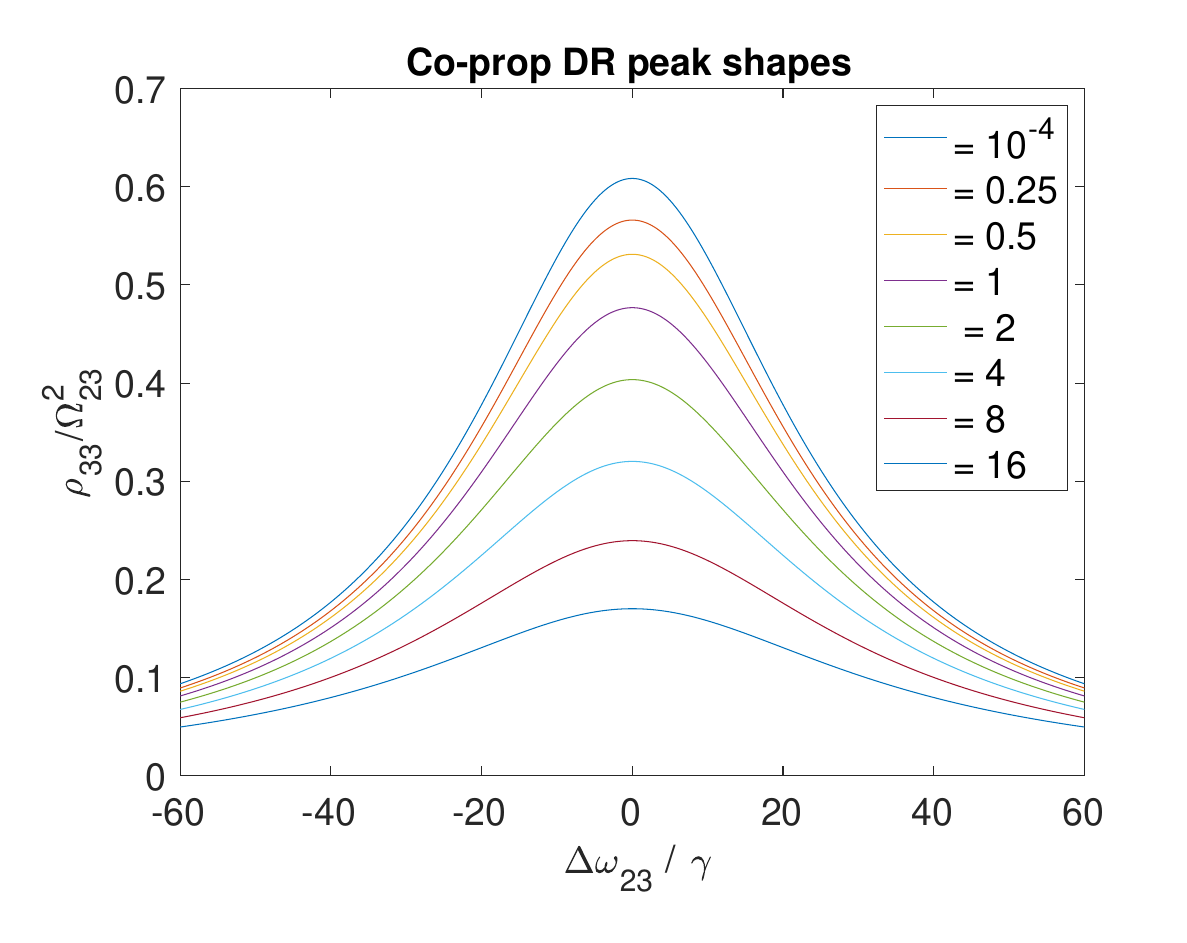}  
\caption{Probe Saturation of  Doppler-averaged Ladder-type Double Resonance with $\Delta \omega_{12} = 0, \omega_b = 2 \omega_a$, and co-propagation of pump.
The legend gives the values of $(\Omega_{23}/\gamma)^2$ values being plotted.   This plot demonstrates that the probe absorption is saturated by a factor of 2
for $(\Omega_{23}/\gamma)^2 \approx 4$}
\label{fig;Probe_saturation_R12=10_co}
\end{center}
\end{figure}

\begin{figure}[h]
\begin{center}
\includegraphics[width=10cm]{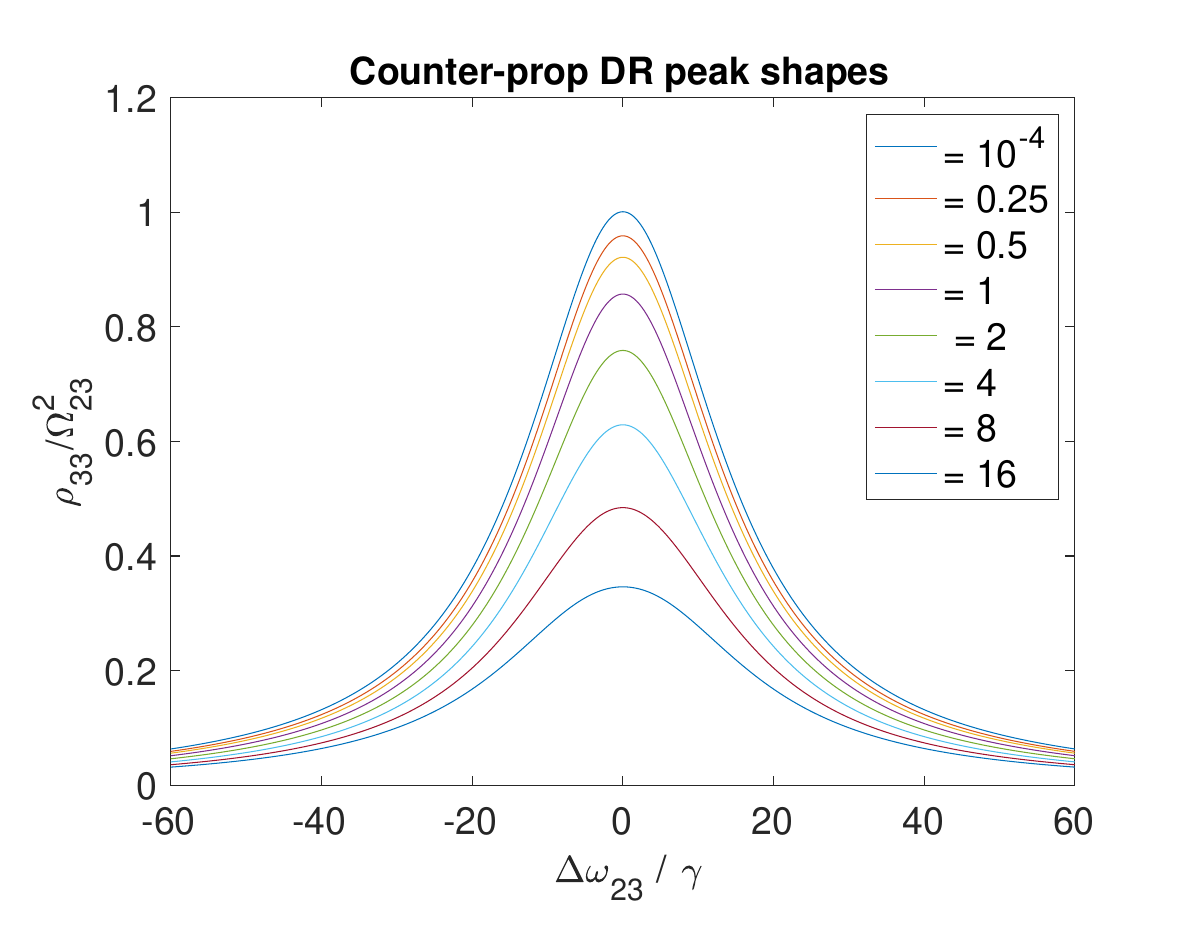}  
\caption{Probe Saturation of  Doppler-averaged Ladder-type Double Resonance with $\Delta \omega_{12} = 0, \omega_b = 2 \omega_a$, and counter-propagation of pump.
The legend gives the values of $(\Omega_{23}/\gamma)^2$ values being plotted.   This plot demonstrates that the probe absorption is saturated by a factor of 2
for $(\Omega_{23}/\gamma)^2 \approx 4$}
\label{fig;Probe_saturation_R12=10_counter}
\end{center}
\end{figure}

  Most significantly, this pump power broadening of the probe transition is a consequence of inhomogeneous broadening.   This is demonstrated by examining the saturation of the double resonance peaks as a function of
  probe intensity.  Figures~\ref{fig;Probe_saturation_R12=10_co} and \ref{fig;Probe_saturation_R12=10_counter} display the evolution of the probe DR spectrum with increasing values of $(\Omega_{23}/\gamma)^2$
  for both co- and counter probe propagation.  These were evaluated with $\Omega_{12} = 10$ and $k_b = 2 k_a$.   In both cases, the probe fractional absorption is reduced by a factor of 2, which
  corresponds to saturation parameter = 1 for a homogeneously broadened line) when $(\Omega_{23}/\gamma)^2 \approx 4$, a factor of 4 higher than for the probe transition without the pump.   If we interpret the enhanced width
  of the DR due to the power broadening by the pump as a homogeneous width of the probe transition, the saturation intensity would be predicted to increase by the square of the ratio of HWHM's of the transition with and  without broadening, {\it i.e.}~657 and 226 times for co- and counter- respectively.   These calculations have only considered absorption by traveling probe waves.     
 
 \clearpage

 \subsection{Double Resonance with standing wave probe}
 The experiments we have performed~\cite{Foltynowicz21a, Foltynowicz21b, deOliveira24, Hjalten24, Lehmann25} used a linear optical cavity, finesse $\approx 1000$, for the probe wave to enhance its absorption and thus molecules were simultaneously excited by both co- and counter-propagating waves.  If $\Delta \omega_{12}$ is sufficiently large that the co- and counter-propagating probe wave DR peaks are well resolved, a given $v_z$ class will strongly interact with at most one of the waves and the standing probe wave DR spectrum is simply the sum of the DR peaks for both directions given above.  Even when the peaks overlap, in the weak probe limit the expected DR spectrum is simply the sum of the co- and counter-propagating spectra.   However,  for even a weakly saturating probe, we need to consider the interaction of the two probe waves.  
 
 Returning to the Dressed-State picture,\cite{Berman11} molecules with fixed $v_z$ will be in resonance for one of the dressed state probe transitions when
 \begin{equation}
\Delta\omega_{23} = k_b v_z + (1/2) \left[  \Delta\omega_{12} + k_a v_z  \pm \sqrt{ \left(\Delta\omega_{12} + k_a v_z \right)^2 + \Omega_{12}^2   }   \right]
\end{equation}
\sloppy
Thus, we can anticipate Lamb dips due to the simultaneous pumping of one of the doublet lines from both directions in the cavity for $v_z \approx 0$ at $\Delta\omega_{23}$
detuning of $(1/2) \left[  \Delta\omega_{12}  \pm \sqrt{ \Delta\omega_{12}^2 + \Omega_{12}^2   }   \right]$.  In addition, we can anticipate  a cross-over resonance due to absorption from one dressed state for the co-propagating probe wave and the
absorption from the other dressed state by the counter-propagating wave.   This will occur when the $\mp k_b v_z$ terms cancel the $\pm(1/2) \sqrt{ \left(\Delta\omega_{12} + k_a v_z \right)^2 + \Omega_{12}^2   }$.
For $\Delta\omega_{12} = 0$, these cross-over resonances occurs when $(4 k_b^2 - k_a^2) v_z ^2 = \Omega_{12}^2$ which
occurs at probe detuning $\Delta\omega_{23} = \pm \frac{k_a}{2 \sqrt{ 4 k_b^2 - k_a^2  }} \Omega_{12} \rightarrow \pm \Omega_{12} / \sqrt{60}$ for $k_b = 2 k_a$.

We have calculated the DR spectrum with a saturating standing wave probe field by numerical integration of the time dependent Schr{\"o}dinger equation in the field 
interaction representation.\cite{Berman11}   
With the same assumptions
and definitions as above,  the time evolution for the amplitudes for the three states with a standing wave probe can be written as
\begin{eqnarray}
\frac{dc_1}{dt} &=& - i \frac{\Omega_{12}}{2} \exp\left( i( \Delta\omega_{12} + k_a v_z)t  \right) c_2  -(\gamma/2) c_1    \nonumber\\
\frac{dc_2}{dt} &=& - i \frac{\Omega_{12}}{2} \exp\left( -i ( \Delta\omega_{12} + k_a v_z)t  \right) c_1   - i \Omega_{23} \exp\left( i \Delta\omega_{23} t  \right) \sin(k_b v_z t + \phi) c_3   -(\gamma/2) c_2   \nonumber\\
\frac{dc_3}{dt} &=&  - i \Omega_{23} \exp\left( -i \Delta\omega_{23} t  \right) \sin(k_b v_z t + \phi) c_2  -(\gamma/2) c_3  
\end{eqnarray}
with $\phi = k_b z(0)$ and $c_n$ is the amplitude of state $\left|n\right>$.
The rate of collisional production of molecules per unit volume is $\gamma N_1$ where $N_1$ is the equilibrium number density of molecules in state $1$.
Assuming the initial conditions at $t=0$ are $c_i = \delta_{i1}$ we can evaluate the steady-state density matrix elements as $\rho_{ij}(v_z) = \gamma N_1 \int_0^{\infty} c_i^*(t) c_j(t) dt$.
The differential equations for $c_i$ and $\rho_{33}$ were numerically evaluated using Matlab's ode45 integrator.   Integration over $k_a v_z$ was then carried out over the interval between $\pm200\gamma$ with
step size of $0.2\gamma$, repeated for $\phi =  [1,3 \ldots 9] \pi/20$.   Calculations were performed for $\Delta\omega_{12} = 0$ and $k_b = 2 k_a$ and a range of $\Delta\omega_{23}$.
Due to the computational cost of solving the differential equations, the evaluation of the DR line shape is much more time consuming, taking about 1 min per $\Delta\omega_{23}$ value on
a recent Mac Laptop.  

\begin{figure}[h]
\begin{center}
\includegraphics[width=10cm]{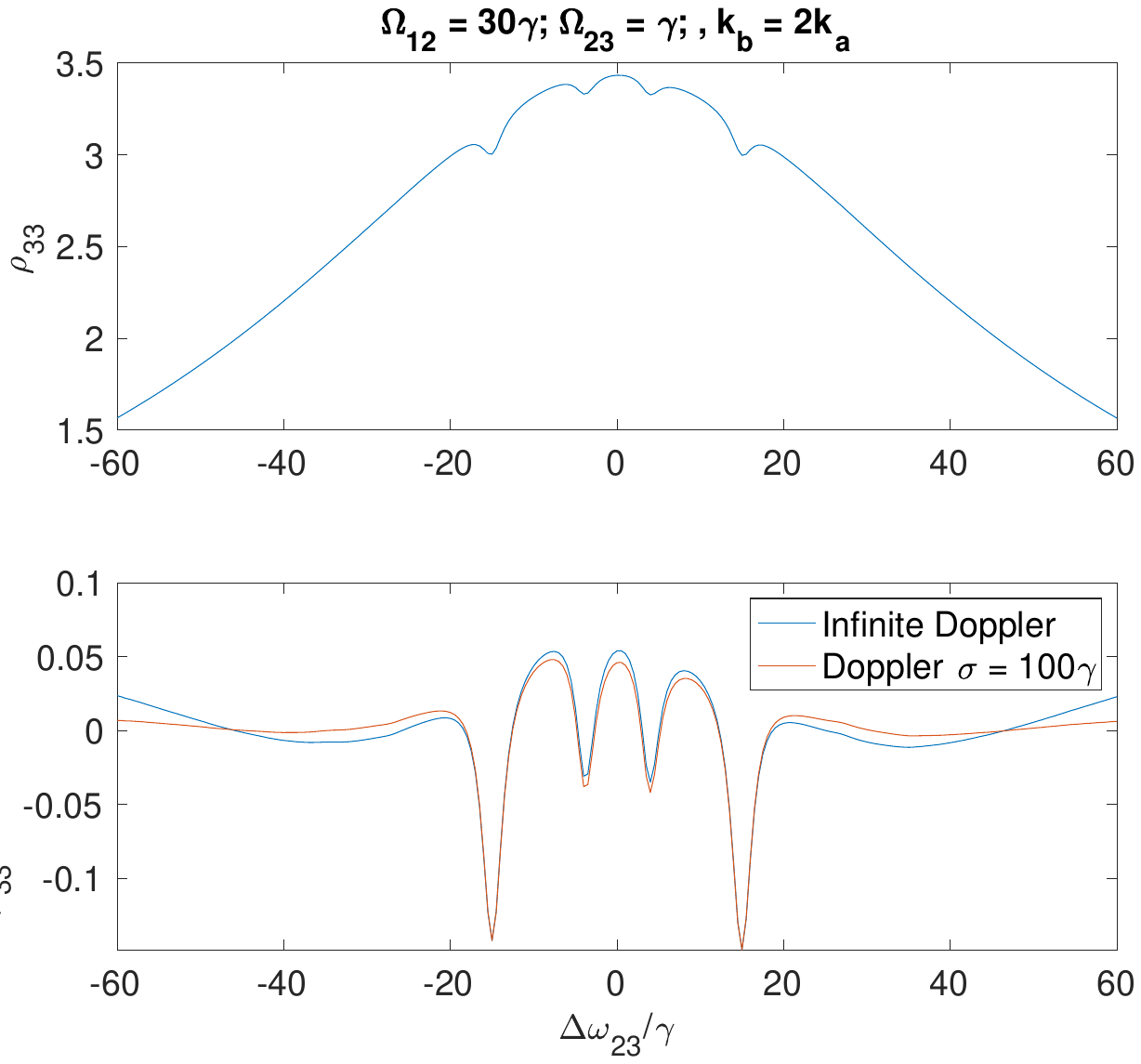}  
\caption{Calculated Double Resonance Spectrum for standing probe wave with $\Omega_{12} = 30 \gamma, \Omega_{12} = \gamma$.  Spectra were integrated
over range $k_a v_z = [-200, +200] \gamma$ with and without Doppler weighting. The Doppler weighting was assumed to have standard deviation $100\gamma$.  
The upper panel shows the spectrum and the lower the difference compared to a Lorentzian fit.  The outer pair of peaks are at $ \Delta\omega_{23} = \pm 15 \gamma$
and inner ones at $\pm 2.6\gamma$ as predicted for the Lamb-Dip and cross-over resonances by the Dress-state model.   }
\label{fig;rho33vsdw23-R12=30_R23=1HR}
\end{center}
\end{figure}

Figure~\ref{fig;rho33vsdw23-R12=30_R23=1HR} presents a plot of the probe absorption spectrum for one such calculation.   $\Omega_{12}$ was taken as $30\gamma$ to 
separate the four predicted peaks and $\Omega_{23} = \gamma$ to produce modest probe saturation.  Displayed is the total spectrum and the difference of the calculated DR
spectrum from a Lorentzian fit to the entire peak.  The calculation was done for both the infinite Doppler width approximation and assuming a
Doppler width with $\sigma = 100 \gamma$ for the pump transition.  There is little difference for the two results.
 Clearly, there are four narrow dips in the DR peaks with widths much narrower than the overall pump power broadened DR
peak, which has HWHM of about $2\Omega_{12}$, the average of the co- and counter-propagating traveling probe wave peaks.
As predicted by the Dressed state picture, the two Lamb dip transitions are centered at $\Delta\omega_{23} = \pm \Omega_{12}/2$.
Weaker, cross-over resonances are centered at $\pm 0.14\Omega_{12}$, again as predicted.  The widths of these features are slightly more than
$\gamma$, consistent with modest probe power broadening.  This result clearly demonstrates that the pump power broadening in inhomogeneous and
is qualitatively different from the power broadening a two level system.

 \section{ V-type Double resonance }
 
 For V-type DR, the equilibrium population in now in state $2$.  The general steady-state solution to the three level density matrix is again given in the Appendix, with
 elements labeld $\rho_{ij}^{\rm V}$.  In the weak prove limit, $\Omega_{23} << \gamma$, the population terms can ca simplified to
  \begin{eqnarray}
 \rho_{ij}^{\rm V,wp} & = & \rho_{ij}^{\rm V,wp*} = \frac{{\rm Num}^{\rm V.wp}_{ij}}{ \rm DEN }   \hspace{1in}  \rho_{22}^{\rm V,wp} = 1 - \rho_{11}^{\rm V.wp}  -\rho_{33}^{\rm V,wp}   \\
{\rm Num}^{\rm V,wp}_{33} &=& \Omega_{23}^2 \left( 16 \gamma^4+ 16 \Delta\omega_{21}^4 - 32 \Delta\omega_{21}^3 \Delta\omega_{23} +
 16 \Delta\omega_{21}^2 \Delta\omega_{23}^2 \right. \nonumber \\
 && + \left(16 \Delta\omega_{21}^2    -28 \Delta\omega_{21} \Delta\omega_{23} +   12 \Delta\omega_{23}^2 \right) \Omega_{12}^2  
   +\Omega_{12}^4   \nonumber \\
 &&  \left. + 8  \gamma^2 ( 4 \Delta\omega_{21}^2 - 4 \Delta\omega_{21} \Delta\omega_{23} + 2 \Delta\omega_{23}^2 + \Omega_{12}^2 )  \right)  \\
 {\rm Num}^{\rm V,wp}_{11} &=& \Omega_{12}^2 \left( 16 \gamma^4+ 16 \Delta\omega_{23}^4 - 32 \Delta\omega_{23}^3 \Delta\omega_{21} \right. \nonumber \\
 && +
16 \Delta\omega_{23}^2 \Delta\omega_{21}^2 + \left(16 \Delta\omega_{23}^2    -28 \Delta\omega_{23} \Delta\omega_{21} +   12 \Delta\omega_{21}^2 \right) \Omega_{23}^2  \nonumber \\
&&  + 8 \Delta\omega_{23} (\Delta\omega_{21} +\Delta\omega_{23}  )\Omega_{12}^2   +\left(\Omega_{12}^2+\Omega_{23}^2 \right)^2   \\
 && \left. + 8  \gamma^2 ( 4 \Delta\omega_{23}^2 - 4 \Delta\omega_{23} \Delta\omega_{21} + 2 \Delta\omega_{21}^2 + \Omega_{12}^2 + \Omega_{23}^2 )  \right) \nonumber \\
\end{eqnarray}  
where DEN is the same as for the Ladder-type solution give above in eq\,\ref{eq:DENwp} but with $\Delta\omega_{12}$ replaced by $-\Delta\omega_{21}$.   It is noted
that the expressions for $\rho_{11}^{\rm V,wp} $ and $\rho_{33,wp}^{\rm V} $ are symmetric under the exchange $ 1 \leftrightarrow 3$ in the subscripts.  
The case where the initial population is entirely in
state $3$ can be calculated as the identify matrix minus the density matrices for initial population in states $1$ and $2$, 
or by inverting the $1$ and $3$ labels in the ladder-type DR expressions.  The general case, with thermal population in all three levels, can be written as the sum of these three
density matrices weighted by the fractional equilibrium population in each of the three states.    

In the limit of weak pump and probe, the expression for $\rho^{\rm V,wp}_{33}$ further simplifies to
\begin{eqnarray}
&&\rho^{\rm V, wpp}_{33} = \frac{ \Omega_{23}^2}{ 2( \gamma^2 + \Delta\omega_{23}^2 + \Omega_{23}^2    )}  \label{eq:rhoVwpp33} \\
&& - \frac{ \left( 4 \gamma^4 + \gamma^2 (2 \Delta\omega_{21}^2 + \Delta\omega_{21} \Delta\omega_{23} + \Delta\omega_{23}^2) + \Delta\omega_{23}(\Delta\omega_{21}-\Delta\omega_{23})( 2 \Delta\omega_{21}^2 - \Delta\omega_{23}^2 )   \right) \Omega_{12}^2 \Omega_{23}^2}
{ 8 (\gamma^2 +\Delta\omega_{23}^2)^2(\gamma^2 + \Delta\omega_{21}^2)(\gamma^2 + (\Delta\omega_{21}-\Delta\omega_{23})^2)} \nonumber 
\end{eqnarray}
The first term on the right is the background probe absorption and the 2nd the DR dip in the probe absorption with resonances at $\Delta\omega_{12} = 0$, $\Delta\omega_{23} = 0$,
and $\Delta\omega_{12} = \Delta\omega_{23}$.  The last is the resonance for the Raman transition between states 1 and 3.
The expression for $\rho_{11}^{\rm V, wpp}$, which is proportional to the pump wave absorption, is the same as Eq.~\ref{eq:rhoVwpp33}
if one swaps the $1,3$ labels.
Making the substitution to account for Doppler shifts and integration over $v_z$ gives a Lorentzian dip in the background probe absorption with HWHM
equal to $\gamma (1 + k_b/k_a)$ for the counter-propagating probe.  For co-propagating probe, the HWHM equals $\gamma$ for 
$k_b < k_a$ and $ k_b \gamma$ for $k_b > k_a$.  This is like for the Ladder type DR but with the co- and counter-propagation values
inverted.  The expression for $\rho_{11}^{\rm V, wpp}$, which is proportional to the weak pump wave absorption, is the same as Eq.~\ref{eq:rhoVwpp33}
if one swaps the $1,3$ labels.

\begin{figure}[h]
\begin{center}
\includegraphics[width=10cm]{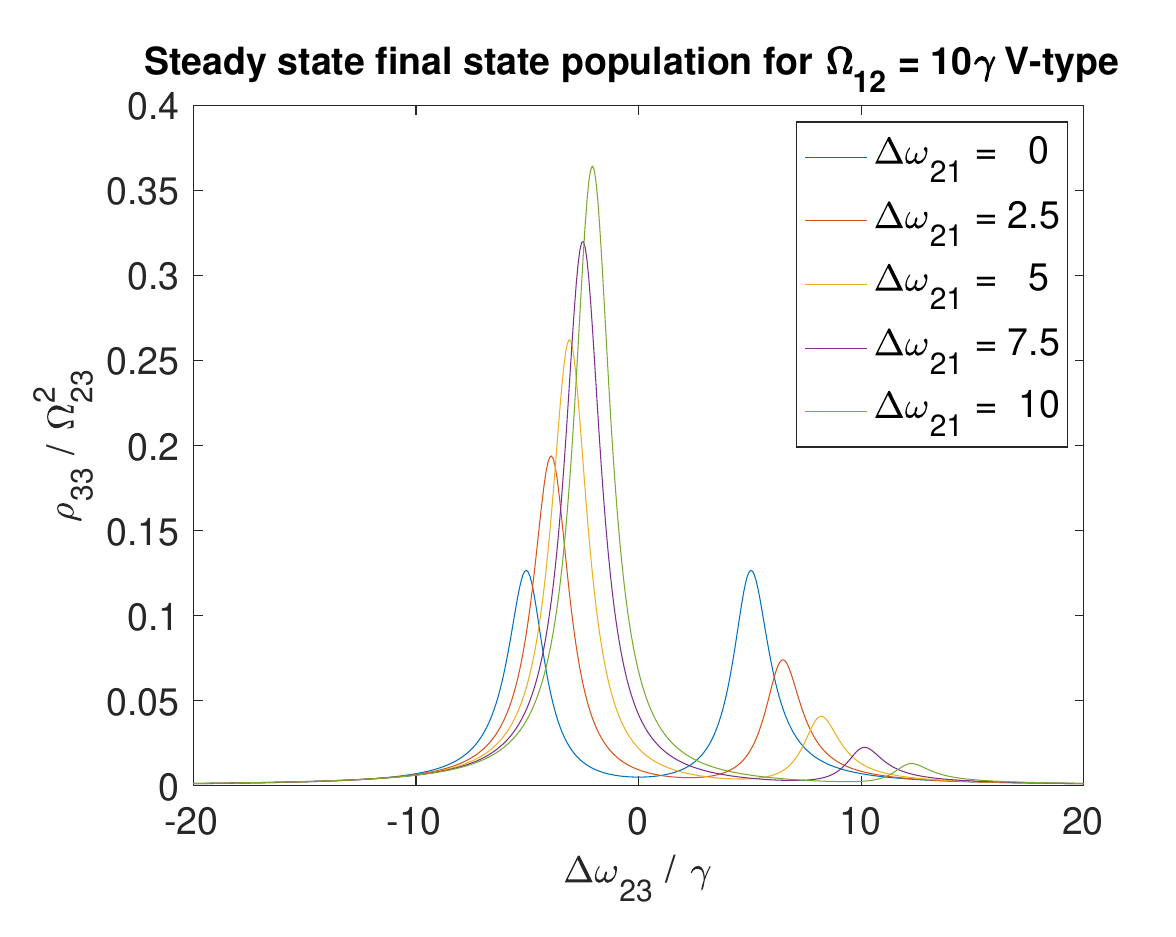}  
\caption{V-type weak probe Double resonance probe absorption spectra for several values of $\Delta\omega_{21}$, in units of $\gamma$.}
\label{fig:V_type_No_Doppler}
\end{center}
\end{figure}

Figure~\ref{fig:V_type_No_Doppler} displays the calculated probe absorption spectrum  in the Doppler-Free case for several values of $\Delta\omega_{21}$.  Like for the ladder type, the spectrum for $\Delta\omega_{21} = 0$ is a pair of near Lorentzian peaks with HWHM equal to $\gamma$ and centers shifted to $\pm |\Omega_{12}|/2$.  This is as
expected as these are the predicted Dressed State resonances and each dressed state is a 50-50\% mixture of states 1 and 2.  As $\Delta\omega_{21}$ grows, one
peak shifts towards the $2 \leftrightarrow 3$ resonances and the other to the $1 \leftrightarrow 3$ Raman type resonance.  However, rather than the peaks
having equal heights, as in the ladder type case, the peak moving towards $\Delta\omega_{23} = 0$ takes most of the intensity as now the the dressed state
moving to $\Delta\omega_{23} = 0$ in the $\Delta\omega_{21}^2 >> \Omega_{12}^2$ limit is almost entire state $2$

The case of Doppler broadened pump and probe lines is treated by replacing $\Delta \omega_{12}$ by 
$\Delta \omega_{12} + k_a v_z$ and $\Delta \omega_{23}$ by $\Delta \omega_{23} \pm k_b v_z$ in
the expressions for the elements of $\rho^{\rm V}$ and then integrating over $v_z$.  
As for the ladder-type DR case, Mathematica was not able to produce analytical results for integration over $v_z$ except for the weak pump and probe limit.
In that limit, $\rho_{33}^{\rm V}$ equals to the steady-state value for the probe field, which is proportional to the Doppler broadened probe absorption line, and
a negative Lorentzian DR lineshape of HWHM = $\gamma$, which reduces that background absorption.  

\begin{figure}[h]
\begin{center}
\includegraphics[width=15cm]{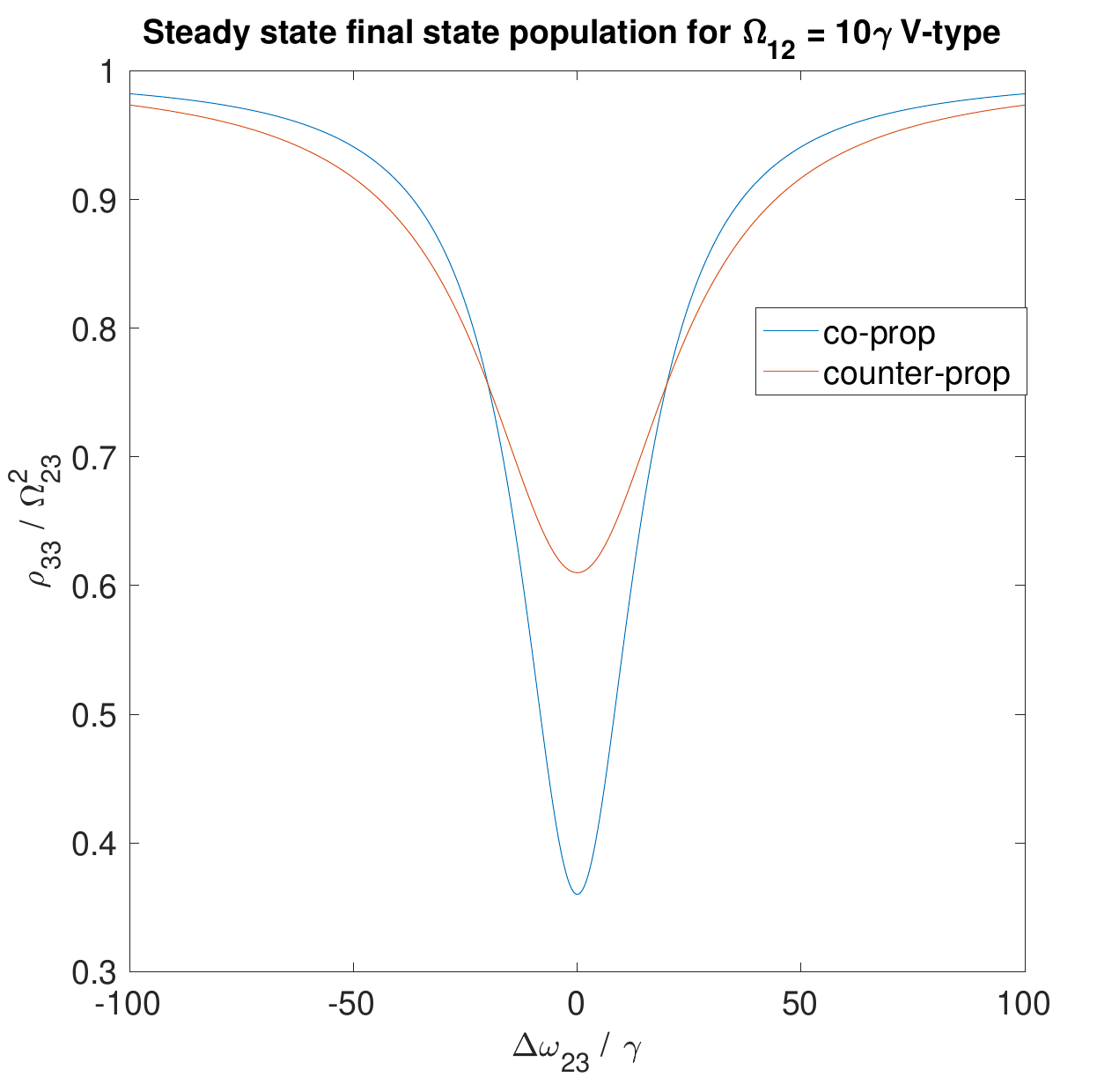}  
\caption{Doppler convoluted  V-type weak-probe Double resonance probe absorption spectra for $\Omega_{12} = 10 \gamma$, $\Delta\omega_{12} = 0$.  The probe absorption has been normalized to its
strength without the pump field.}
\label{fig:V-type_Doppler_R12=10}
\end{center}
\end{figure}

We have numerically integrated $\rho_{33}^{\rm V}$ over $v_z$ to treat the cases with pump saturation for a range of values of $\Omega_{12}$ with $k_b = 2 k_a$ and also for a range of values
of $k_b / k_a$ values with $\Omega_{12} = 10\gamma$.
Figure~\ref{fig:V-type_Doppler_R12=10} displays the calculated probe absorption spectrum for V-type DR after convolution over an infinite Doppler width in the weak probe limit
for $k_b = 2k_a$ and $\Omega_{12} = 10\gamma$.  The spectra have been normalized so
the probe absorption without pump equals 1.   Here, the DR signal appears as a dip in the probe absorption that can be viewed as due to depletion of the center of the Doppler broadened absorption due to the
Bennett hole\cite{Bennett62} in the lower (2) state population produced by the pump wave.  Again, in the limit that $\Omega_{12} >> \gamma$, and $\omega_b = 2 \omega_a$, the depletion of the probe 
absorption has a Lorentzian lineshape with equal areas but with HWHM approximately equal to 1.5 and 2.5 times $\Omega_{12}$.  Note, however, that the absorption of the co-propagating wave gives the narrower and stronger depletion, opposite to
the ladder-type considered above.  For $\Omega_{12} = 10\gamma$, the HWHM of the fitted Lorentzian lines are 15.575 and 25.626$\gamma$, identical but flipped with respect to directions, as the Ladder-type DR positive peaks.  Calculations for different
values of $k_b / k_a$ between 1-4 gives 
HWHM values equal to  $( k_b / k_a \mp 0.5 ) \Omega_{12}$ for co- and counter-propagation.   For $ k_b / k_a < 1$
the co-propagation peak is no longer Lorentzian.  

\clearpage

\begin{figure}[h]
\begin{center}
\includegraphics[width=15cm]{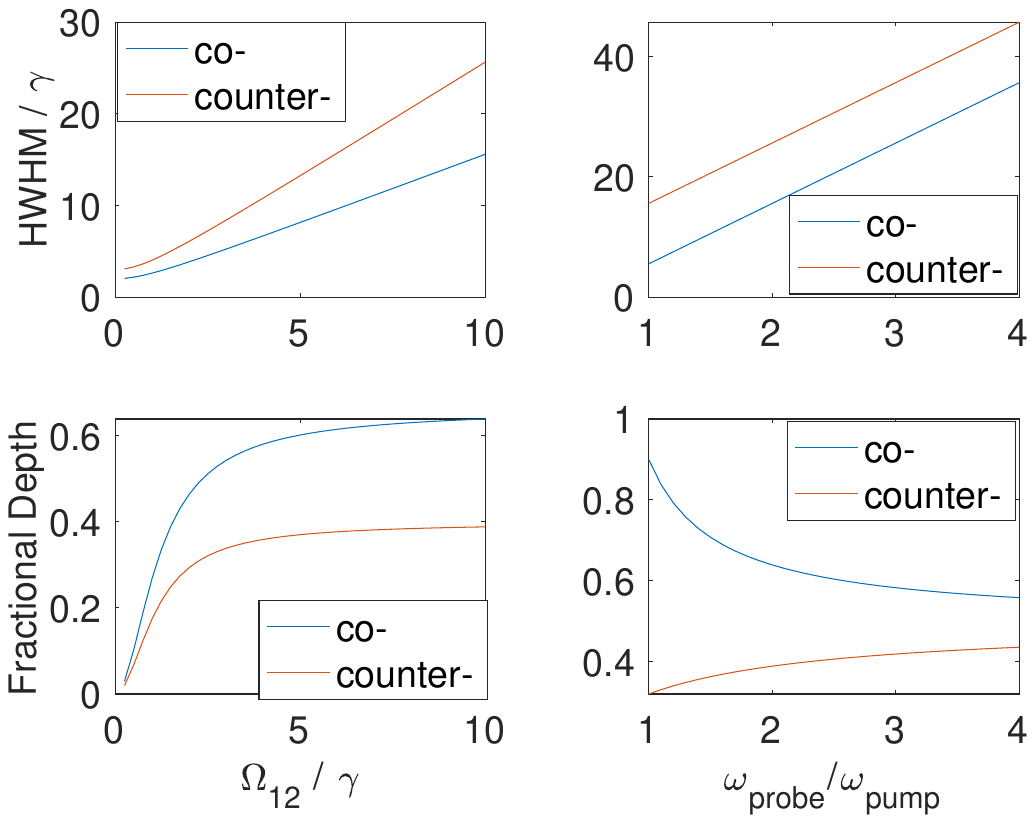}  
\caption{Lorentzian peak parameters for weak probe, Doppler convoluted V-type Double resonance probe absorption spectra for $\Omega_{12} = 10 \gamma$}
\label{fig:V-type_Doppler_peaks}
\end{center}
\end{figure}

Figure~\ref{fig:V-type_Doppler_peaks}, displays the Lorentzian peak fit parameters from fits of the calculated VDR dip lineshapes calculated for $\Omega_{12} = 0.25-10.\gamma$ for $k_b = 2k_a $
and for calculations with $k_b / k_a = 1-4$ with $\Omega_{12} = 10\gamma$.  The HWHM values are nearly identical to those of the LDR case but with co- and counter-propagation
results flipped.  For larger values of $k_b / k_a$, the fractional depth of the VDR dip approaches 0.5 (from opposite sides for the two propagation directions).  This is the dip depth
expected from a simple rate equation model where the strong pump would be expected to lower the equilibrium population in state 2 by a factor of 1/2 in the strong pump saturation limit.

 \clearpage
 
 \subsection{ Comparison with Experiment}
 We will provide a single example of comparison with experiment, in this case Double resonance pumping the CH$_4$ $\nu_3$ R(0) line and probing the $2\nu_3$ R(0) line.\cite{deOliveira25}
 The experimental conditions were a sample pressure of 11 mtorr and a pump power of 41 mW.  Using the Lamb-dip pressure broadening coefficient.\cite{Barger69}, we estimate the relaxation rate to be $\gamma =$ 1.1E6/sec.   The beam radius was 7.1\,mm., thus the on axis pump intensity of 5.2W/cm$^2$.  The pump transition has transition dipole moment of 0.093 Debye, which gives a Rabi frequency $\Omega_{12} =$ 8.7E6/sec = 7.9\,$\gamma$.  The ratio of probe over pump frequencies is close to 2.0.  The pump was detuned by 3 MHz from the center of the line, which yielded a pair of V-type DR peaks separated by the expected 12 MHz which were well fit to a pair of Lorentzians with HWHM of 3.324(5) (co-propagation) and 1.869(3) (counter propagation), both far below the 230 MHz Doppler HWHM of the probe transition.  The model of this paper predicts widths of 3.49 and 2.09 MHz respectively, slightly too high.  Likely, part of the discrepancy is that the probe (which has a radius $\sqrt{2}$ smaller than the pump) samples molecules away from the axis so the effective Rabi frequency is overestimated).  Averaging  the radial Rabi frequency over the radial intensity of the probe gives an effective Rabi frequency 0.89 times the on axis value, and reduces the predicted widths to 3.10 and 1.86 MHz, in excellent agreement with the experimental values.  We have experimentally verified that the width of the observed DR features scales as the square root of the pump intensity, as predicted by the model.

 \section{Inverted pump-probe}
 In a recent experiment, we observed $\Lambda$-type DR with the sequence g.s. $\leftrightarrow 2\nu_3 \leftrightarrow \nu_3$ with the "strong" field driving the transition between the pair of 
 states without thermal population.  We monitor the change in absorption of the NIR $\omega_{12}$ wave as the intensity of the saturating MIR $\omega_{23}$ was amplitude modulated.  We can calculate the average
 rate of NIR absorption as $\gamma( \rho_{22} + \rho_{33}) $; there must have been one NIR absorbed photon each time a molecule relaxes from states $2$ or $3$.    We use the Ladder
 DR expression but with the signs of $\Delta\omega_{23}$ and $k_{23}$ switched as the transition $2 \rightarrow 3$ results in stimulated emission instead of absorption.  
 
Without Doppler broadening, turning on $\Omega_{23}$ results in a splitting of $\omega_{12}$ transition due to the Autler-Townes effect. When $\Delta\omega_{23} = 0$, the NIR transition,
$\omega_{12}$,
is split into two equal height peaks shifted by $\pm \Omega_{23}/2$ each with HWHM=$\gamma$.  Thus when $\Omega_{23} >> \gamma$, we should expect essentially 100\%
modulation of the NIR absorption with the pump depleting the on-resonance ($\Delta\omega_{12} = 0$) absorption.

\begin{figure}[h]
\begin{center}
\includegraphics[width=17cm]{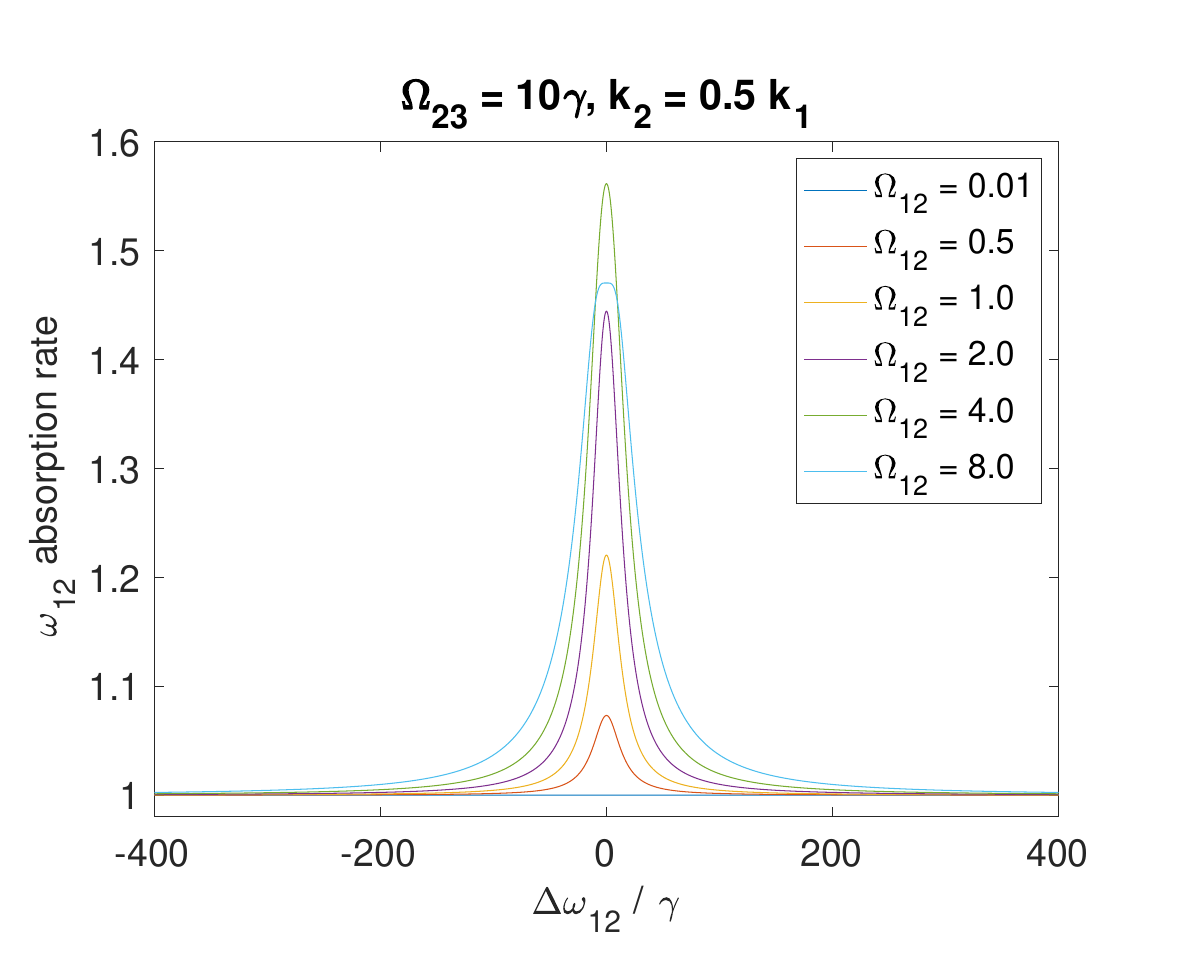}  
\caption{Calculated Doppler convoluted  $\omega_{12}$ absorption for $\Omega_{23} = 10\gamma$ divided by that
when $\Omega_{23} = 0$ for co-propagation and several values of $\Omega_{12}/\gamma$ and  $\Delta\omega_{23} = 0$.  Note that at least modest $\omega_{12}$ saturation is required to observe the DR feature, and that
the DR peak results in increased absorption.}
\label{fig:Inverted_pump_probe_Doppler_p}
\end{center}
\end{figure}

\begin{figure}[h]
\begin{center}
\includegraphics[width=17cm]{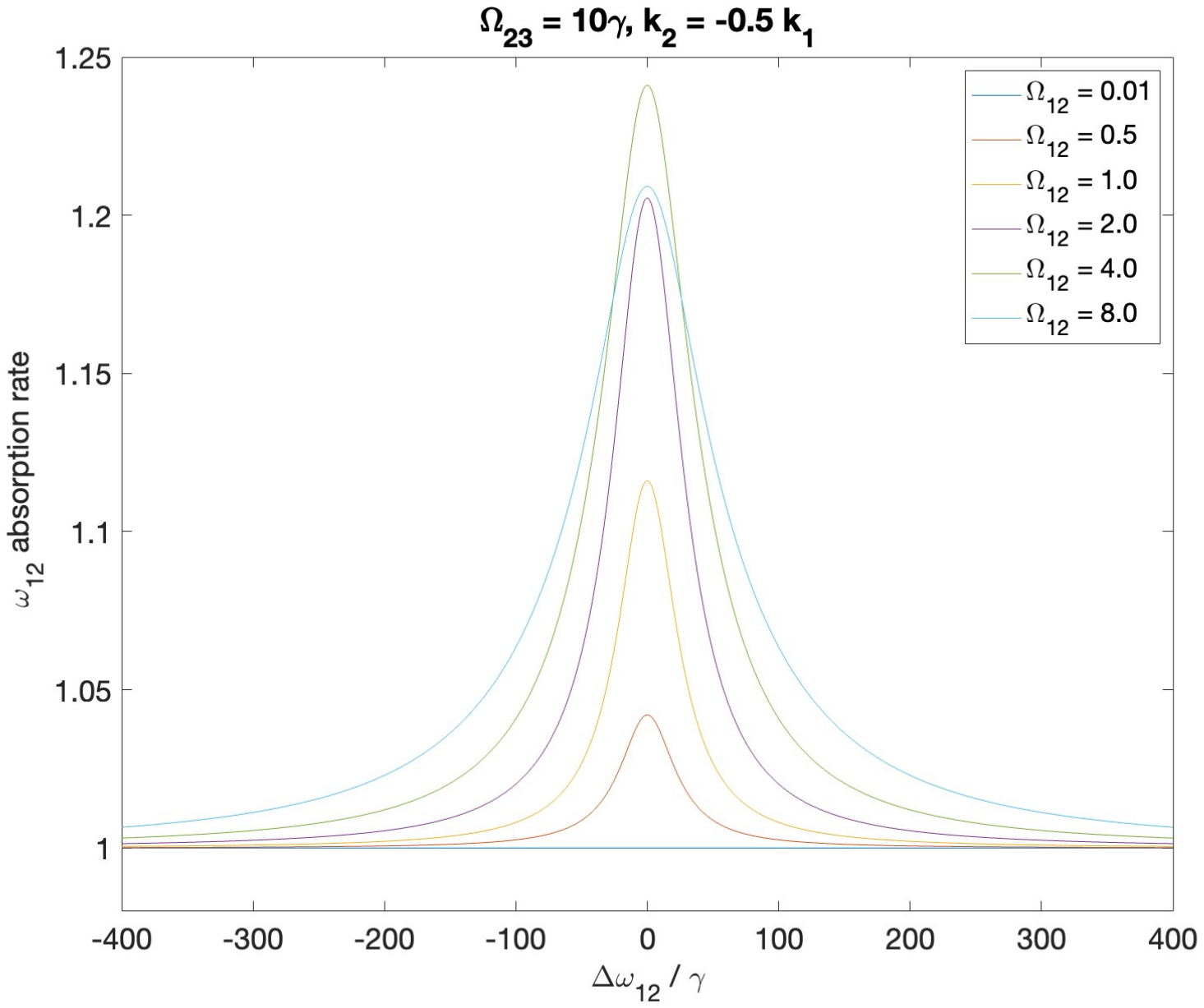}  
\caption{Calculated Doppler convoluted  $\omega_{12}$ absorption for $\Omega_{23} = 10\gamma$ divided by that
when $\Omega_{23} = 0$ for counter-propagation and several values of $\Omega_{12}/\gamma$ and  $\Delta\omega_{23} = 0$.  Note that at least modest $\omega_{12}$ saturation is required to observe the DR feature, and that
the DR peak results in increased absorption.}
\label{fig:Inverted_pump_probe_Doppler_m}
\end{center}
\end{figure}

\begin{figure}[h]
\begin{center}
\includegraphics[width=17cm]{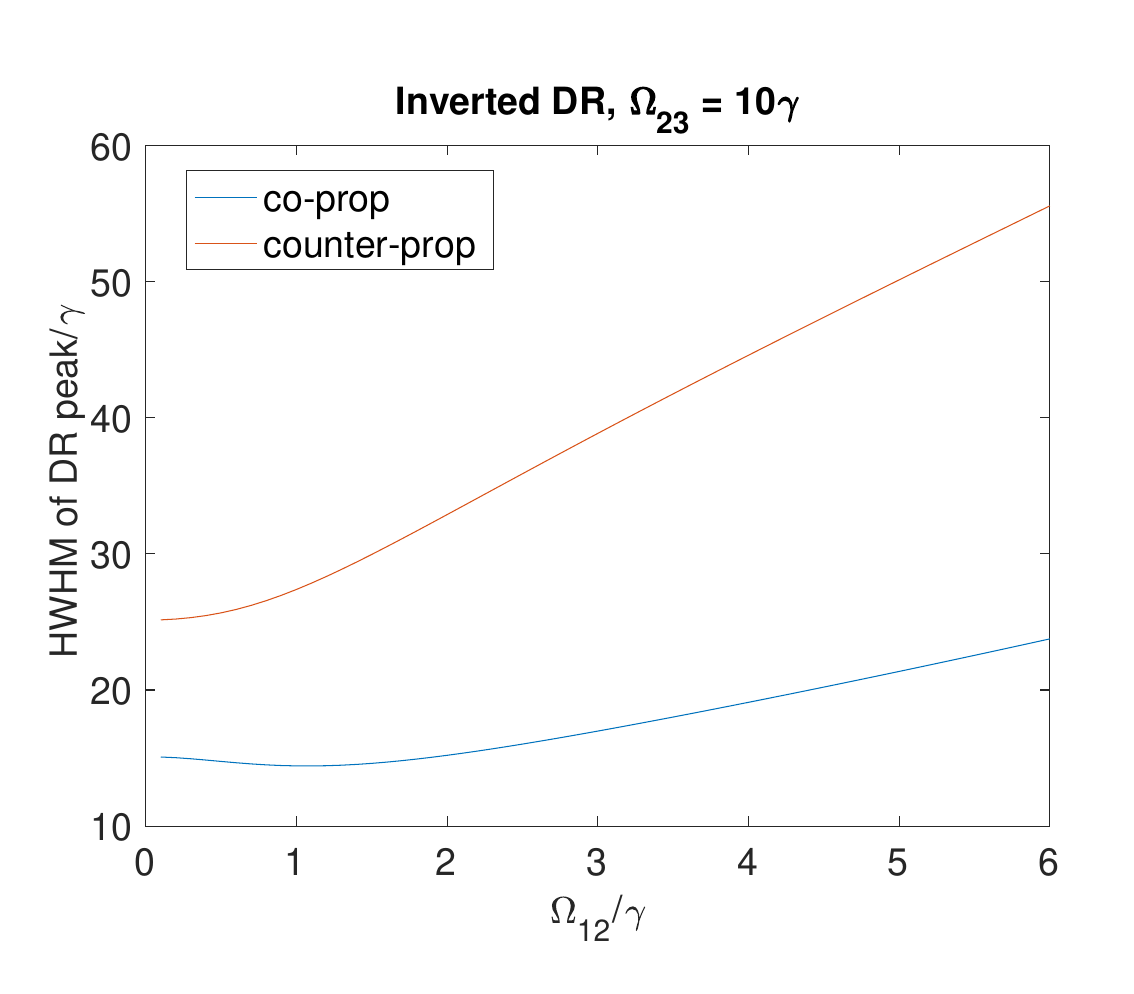}  
\caption{HWHM of Lorentzian Fit to DR absorption enhancement as a function of $\omega_{12}$ for (dump) $\Omega_{23} = 10\gamma$.}
\label{fig:Inverted_width_vs_R12}
\end{center}
\end{figure}

\begin{figure}[h]
\begin{center}
\includegraphics[width=17cm]{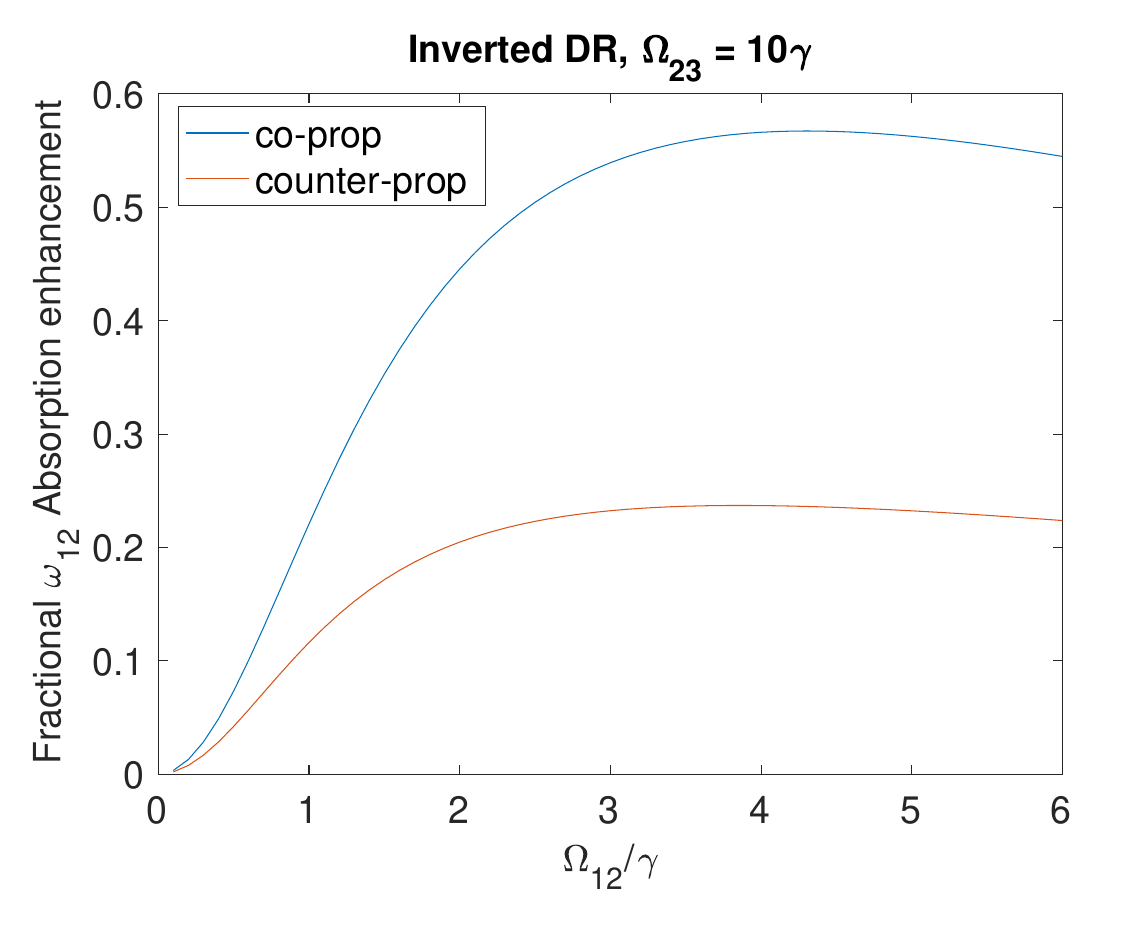}  
\caption{Fractional peak of Lorentzian Fit to DR absorption enhancement as a function of $\omega_{12}$ for (dump) $\Omega_{23} = 10\gamma$.}
\label{fig:Inverted_depth_vs_R12}
\end{center}
\end{figure}

However, in the case of Doppler width $>> \Omega_{23}$ and $\Omega_{12} << \gamma$, after convolution over $v_z$ the calculated DR signal has no modulation of the NIR absorption when the MIR laser is
amplitude or frequency modulated.   Figures~\ref{fig:Inverted_pump_probe_Doppler_p} and \ref{fig:Inverted_pump_probe_Doppler_m} display the calculated NIR absorption for waves co- and counter-propagating respectively, as a function $\Delta \omega_{12}$ for several values of $\Omega_{12}/\gamma$ when
$\Omega_{23} = 10 \gamma$.  In each case, the absorption has been normalized to the absorption for the same value of $\Omega_{12}$ but with $\Omega_{23} = 0$.   The fractional modulation
of the absorption is only about 5\% for $\Omega_{12} = 0.5\gamma$.   Unlike the case without Doppler broadening, the presence of the MIR field increases the absorption of the NIR wave.  The MIR is
depleting the population in state 2 and thereby reduces the $ 2 \rightarrow 1$ stimulated emission at the $\omega_a$ frequency.   Figures~\ref{fig:Inverted_width_vs_R12} and \ref{fig:Inverted_depth_vs_R12} are plots of the
DR HWHM and peak fractional enhancement for Lorentzian fits to the $\omega_{12}$ absorption as a function of $\Omega_{12}/\gamma$ for $\Omega_{23} = 10\gamma$
and $k_a = 2 k_b$.  
The counter propagating DR is weaker and broader, again approximately
in the ratio of 3:5 for the NIR frequency twice the MIR frequency for low $\Omega_{12}$ but rises to 3:7 for $\Omega_{12}  = 6\gamma$.  The areas of the co- and counter-propagating
DR peaks remain close to equal.  Surprisingly, the width of the co-propagating DR peak initially decreases with $\Omega_{12}$, reaching a minimum for $\Omega_{12} \approx \gamma$.
If the strong ``dump'' field is detuned by $\Delta\omega_{23}$ from the center of the $2 \rightarrow 3$ line, the DR peaks as a function of $\Delta\omega_{12}$  will split into a pair of Lorentzian lines
centered on detuning values $\Delta\omega_{12} = \mp (k_a/k_b) \Delta\omega_{23}$ with upper (lower) sign for co- and counter-propagation directions respectively.

It is noted that when we set-up this $\Lambda$-type DR experiment, we anticipated that the strong ``dump'' field would produce an Electromagnetically induced transparency (EIT) effect of dramatically reducing the pump absorption.\cite{Boller91}  
However, as discussed by Fleischhauer et al.,~\cite{Fleischhauer05} strong EIT requires the $\rho_{13}$ decoherence rate to be slow compared to the $\rho_{22}$ relaxation rate, while in our model, all the relaxation rates are equal.   Also, the EIT is largely washed out after Doppler averaging unless the coupling field Rabi frequency, $\Omega_{23}$ in the present
case, exceeds the Doppler width for the 1-3 transition,~\cite{Fleischhauer05} which in turn is on the order of $|k_a \pm k_b|$.  Thus, no EIT is expected for the infinite Doppler width limit.  In the experiments we performed, the residual Doppler broadening, even for the narrower co-propagation case, exceeded $\Omega_{23}$ by more than an order of magnitude.

 \clearpage

\section{ Summary and Conclusions}

In this paper, we have presented a theoretical analysis of the lineshapes expected for OODR transitions under conditions of monochromatic pump and probe
waves and homogeneous widths much less than Doppler broadening of the transitions.  We used the analytical steady-state solutions of the 3-level density
matrix that result when all relaxation rates are assumed to be equal, which considerably simplifies the results.  It is found that DR signals have Lorentzian lineshapes
when the probe wavenumber exceeds that of the pump laser.   The high pump power signals have widths proportional to the pump Rabi frequencies with a
scale factor that is linearly dependent on the ratio of probe over pump wavenumber and is different for co- and counter-propagating waves.  This width, however, is
inhomogeneous, resulting from the convolution of the Doppler shifts in the Autler-Townes doublets in the homogeneous probe spectrum produced by the strong
pump.   Probe power saturation is similar to that predicted for the bare probe transition and vastly larger higher than predicted if one interprets the power broadening
as homogeneous.  

The present results allow for simple interpretation of the observed double resonance spectra involving rovibrational transitions in low pressure gases such have
recently been performed in several laboratories. 

\section{Acknowledgements}
The author acknowledges stimulation by and discussions with his University of Ume\r{a}
collaborators Vinicius Silva De Oliveira, Aleksandra Foltynowicz, Adrian Hj\"{a}lt\'{e}n, Andrea Rosina, and Isak Silander.  
He also acknowledges helpful discussions with Patrick Dupr\'{e}.
This work was supported by the US National Science, grant CHE02108458, and the Wenner Gren Foundation, grant number GFOv2024-0010.

 \clearpage

\bibliography{Double_Resonance_Spec}

\section{Appendix}
The steady state solutions to the Optical Bloch equations in the rotating wave approximation, Eqs.~\ref{eq:rho_ss_eq},  are functions of the two resonance detuning values $\Delta\omega_{12} = (E_2 - E_1)/\hbar  -  \omega_a$, 
and $\Delta\omega_{23} = (E_3 - E_2)/\hbar  -  \omega_b$; the two Rabi frequencies $\Omega_{12} =  \, \left<2| \vec \mu |1 \right>\cdot{\cal \vec E}_{a} / \hbar$ and $\Omega_{23} =  \, \left<3| \vec \mu |2 \right>\cdot{\cal \vec E}_{b} / \hbar$ for the pump and probe transitions, and the relaxation rate $\gamma$.   The two photon resonance detuning is given by $\Delta\omega_{13} = \Delta\omega_{23} - \Delta\omega_{12}$.
 Using Mathematica, the steady state solutions to Eqs.~\ref{eq:rho_ss_eq} for ladder-type DR equilibrium equations were found to be:
  \begin{eqnarray}
\rho_{ij} &=&  \rho_{ji}^*  = \frac{\rm NUM_{ij}}{\rm DEN} \hspace{0.5in}   \hspace{0.5in} \rho_{11} = 1 - \rho_{22} - \rho_{33}  \nonumber \\
{\rm NUM_{22}} &=& \Omega_{12}^2 \left( 16 \gamma^4 + \left( -4 \Delta\omega_{23} (\Delta\omega_{12}+\Delta\omega_{23}) + \Omega_{12}^2 \right)^2    \right. \nonumber \\
&&+  8 \gamma^2 \left(  2 \Delta\omega_{12}^2 + 4\Delta\omega_{12}\Delta\omega_{23} + 4 \Delta\omega_{23}^2  + \Omega_{12}^2\right) \nonumber \\
&& \left.   + 2 \left( 2(\Delta\omega_{12}+\Delta\omega_{23}  )(3\Delta\omega_{12}+4\Delta\omega_{23})+ \Omega_{12}^2 + 4 \gamma^2 \right) \Omega_{23}^2 + \Omega_{23}^4 \right) \nonumber \\
{\rm NUM_{33}}  &=& \Omega_{12}^2 \Omega_{23}^2 \left(12 \gamma^2
   +4 \left(\Delta \omega_{12}^2
   +\Delta \omega_{12} \Delta \omega_{23}+
   \Delta \omega_{23}^2\right)
   +3 \left(\Omega_{12}^2+\Omega_{23}^2\right)\right) \nonumber \\
   {\rm NUM_{12}} &=& \Omega_{12} \left(   -16 i \gamma^5 + 16 \gamma^4 \Delta\omega_{12} + \Delta\omega_{12}(-4 \Delta\omega_{23} (\Delta\omega_{12} + \Delta\omega_{23}) + \Omega_{12}^2     )^2     \right. \nonumber \\
  && \left. +  \left(  4 (\Delta\omega_{12} + \Delta\omega_{23}) (4 \Delta\omega_{12}^2 + 4 \Delta\omega_{12} \Delta\omega_{23} - \Delta\omega_{23}^2) + (6 \Delta\omega_{12} + 5 \Delta\omega_{23}) \Omega_{12}^2  \right) \Omega_{23}^2   \right. \nonumber \\
   && -4(\Delta\omega_{12}+\Delta\omega_{23})\Omega_{23}^4 - 4 i \gamma^3 (4 \Delta\omega_{12}^2 + 8 \Delta\omega_{12} \Delta\omega_{23} + 8 \Delta\omega_{23}^2 + 2 \Omega_{12}^2 + 5 \Omega_{23}^2) \nonumber \\
  &&  +4 \gamma^2 ( 2 \Delta\omega_{12} ( 2 \Delta\omega_{12}^2 + 4 \Delta\omega_{12} \Delta\omega_{23} + 4 \Delta\omega_{23}^2 + \Omega_{12}^2  ) + (3 \Delta\omega_{12} - \Delta\omega_{23}) \Omega_{23}^2  ) \nonumber \\
   && -i \gamma (   - 4 \Delta\omega_{23} (\Delta\omega_{12} + \Delta\omega_{23}) + \Omega_{12}^2   )^2  \nonumber \\
   && \left.  -i\gamma( 16 \Delta\omega_{12}^2 + 32 \Delta\omega_{12} \Delta\omega_{23} + 5 ( 4 \Delta\omega_{23}^2 + \Omega_{12}^2    ) ) \Omega_{23}^2 - 4i \gamma \Omega_{23}^4     \right)  \nonumber \\
   {\rm NUM_{13}} &=& \Omega_{12} \Omega_{23} \left(  - 8 \gamma^4 + 8 \Delta\omega_{12}^2 \Delta\omega_{23}^2 + 8 \Delta\omega_{12} \Delta\omega_{23}^3
   -8 i \gamma^3 ( 2 \Delta\omega_{12} + \Delta\omega_{23}   ) - 6 \Delta\omega_{12} \Delta\omega_{23} \Omega_{12}^2 \right.  \nonumber \\
  && - 4 \Delta\omega_{23}^2 \Omega_{12}^2 + \Omega_{12}^4 + ( 8 \Delta\omega_{12}^2 + 8 \Delta\omega_{12} \Delta\omega_{23} - 2 \Delta\omega_{23}^2 - \Omega_{12}^2  ) \Omega_{23}^2  \nonumber \\
   &&   -2 \Omega_{23}^4 + 2 \gamma^2 ( 4 \Delta\omega_{12}^2 + 4 \Delta\omega_{12} \Delta\omega_{23} - 4 \Delta\omega_{23}^2 + \Omega_{12}^2 - 5\Omega_{23}^2    ) \nonumber \\
   && \left. -2i \gamma ( 8 \Delta\omega_{12} \Delta\omega_{23}^2 + 4 \Delta\omega_{23}^3 - \Delta\omega_{12} \Omega_{12}^2 - 5 \Delta\omega_{23} \Omega_{12}^2 + 4 ( 2 \Delta\omega_{12} + \Delta\omega_{23}  ) \Omega_{23}^2 ) \right)  \nonumber \\
   {\rm NUM_{23}} &=& \Omega_{12}^2 \Omega_{23} \left( -12 i \gamma^3 + 8 \gamma^2 (\Delta\omega_{12} + 2 \Delta\omega_{23}) + (  \Delta\omega_{12} + 2 \Delta\omega_{23} )( 4 \Delta\omega_{23} ( \Delta\omega_{12} + \Delta\omega_{23})  - \Omega_{12}^2 ) \right. \nonumber \\
   && \left. + (8 \Delta\omega_{12} + 7 \Delta\omega_{23}) \Omega_{23}^2 - i \gamma ( 4 ( \Delta\omega_{12}^2 + \Delta\omega_{12} \Delta\omega_{23} + \Delta\omega_{23}^2 ) + 3 (\Omega_{12}^2 + \Omega_{23}^2)  )     \right)  \nonumber \\
   {\rm DEN} &=& 
   2 \left(16 \gamma ^6+8 \gamma ^4
   \left(4 \left(\Delta \omega_{12}^2
   +\Delta \omega_{12} \Delta \omega_{23}
   +\Delta \omega_{23}^2\right)
   +3 \left(\Omega_{12}^2+ \Omega_{23}^2 \right)\right) \right. \nonumber \\
   && \left.
   +\gamma ^2 \left(4 \left(\Delta \omega_{12}^2
   +\Delta \omega_{12} \Delta \omega_{23}+\Delta \omega_{23}^2\right)
   +3\left(\Omega_{12}^2+\Omega_{23}^2\right)\right)^2  \right. \nonumber \\
   && \left.
   +\Omega_{23}^4 \left(-8
   \Delta \omega_{12}^2-8 \Delta \omega_{12}
   \Delta \omega_{23}+\Delta \omega_{23}^2
   +3\Omega_{12}^2\right) \right. \nonumber \\
   && \left. +\Omega_{23}^2 \left(2
   \Omega_{12}^2 \left(10 \Delta \omega_{12}^2
   +19 \Delta \omega_{12} \Delta \omega_{23}
   +10\Delta \omega_{23}^2\right) \right. \right.  \nonumber \\
  && \left. \left. +8 \Delta \omega_{12} (\Delta \omega_{12}+\Delta \omega_{23})
   \left(2 \Delta \omega_{12}^2+2 \Delta \omega_{12} \Delta \omega_{23}
   -\Delta \omega_{23}^2\right)+3\Omega_{12}^4\right) \right.  \nonumber \\
   && \left. +\left(\Delta \omega_{12}^2+\Omega_{12}^2\right) \left(\Omega_{12}^2
   -4 \Delta \omega_{23} (\Delta \omega_{12}+\Delta \omega_{23})\right)^2+\Omega_{23}^6\right) \label{eq:DEN}
   \end{eqnarray}   
   These are obviously rather opaque expressions!  It is noted that only even powers of the two Rabi frequencies appear (except for the linear pre-factors on the coherences) and the values are unchanged if both $\Delta \omega_{12}$ and $\Delta \omega_{23}$
 switch signs.  The dimensionless $\rho$ only depends upon the ratios of parameters, so one can normalized all to $\gamma$.  The first order Doppler shifts are accounted for by replacing $\Delta\omega_{12}$ by $\Delta\omega_{12} - k_a v_z$
 and  $\Delta\omega_{23} \mp k_b v_z$ where $v_z$ is the projection of the molecular velocity on the propagation direction of the pump wave (a) and the signs in the second term are taken as (-) if the probe wave co-propagates with the pump and (+) if the probe counter-propagates.  For the case of $\Lambda$ type DR, one replaces $\Delta\omega_{23}$ by $-\Delta\omega_{32}$ and also switch the sign of the $k_b v_z$ term as now this transition is stimulated emission instead of absorption.
 
 The V-type DR case is found to have steady-state solutions
  \begin{eqnarray}
 \rho_{ij}^{\rm V} & = & \rho_{ij}^{\rm V*} = \frac{{\rm Num}^{\rm V}_{ij}}{ \rm DEN }   \hspace{1in}  \rho_{22}^{\rm V} = 1 - \rho_{11}^{\rm V}  -\rho_{33}^{\rm V}   \\
{\rm Num}^{\rm V}_{33} &=& \Omega_{23}^2 \left( 16 \gamma^4+ 16 \Delta\omega_{21}^4 - 32 \Delta\omega_{21}^3 \Delta\omega_{23} +
 16 \Delta\omega_{21}^2 \Delta\omega_{23}^2 \right. \nonumber \\
 && + \left(16 \Delta\omega_{21}^2    -28 \Delta\omega_{21} \Delta\omega_{23} +   12 \Delta\omega_{23}^2 \right) \Omega_{12}^2  
   + 8 \Delta\omega_{21} (\Delta\omega_{23} 
+ \Delta\omega_{21}  )\Omega_{23}^2   +\left(\Omega_{12}^2+\Omega_{23}^2 \right)^2   \nonumber \\
 &&  \left. + 8  \gamma^2 ( 4 \Delta\omega_{21}^2 - 4 \Delta\omega_{21} \Delta\omega_{23} + 2 \Delta\omega_{23}^2 + \Omega_{12}^2 + \Omega_{23}^2 )  \right)  \nonumber \\
 {\rm Num}^{\rm V}_{11} &=& \Omega_{12}^2 \left( 16 \gamma^4+ 16 \Delta\omega_{23}^4 - 32 \Delta\omega_{23}^3 \Delta\omega_{21} \right. \nonumber \\
 && +
16 \Delta\omega_{23}^2 \Delta\omega_{21}^2 + \left(16 \Delta\omega_{23}^2    -28 \Delta\omega_{23} \Delta\omega_{21} +   12 \Delta\omega_{21}^2 \right) \Omega_{23}^2  \nonumber \\
&&  + 8 \Delta\omega_{23} (\Delta\omega_{21} +\Delta\omega_{23}  )\Omega_{12}^2   +\left(\Omega_{12}^2+\Omega_{23}^2 \right)^2  \nonumber  \\
 && \left. + 8  \gamma^2 ( 4 \Delta\omega_{23}^2 - 4 \Delta\omega_{23} \Delta\omega_{21} + 2 \Delta\omega_{21}^2 + \Omega_{12}^2 + \Omega_{23}^2 )  \right) \nonumber \\
 {\rm NUM^V_{12}} &=& \Omega_{12} \left( 16 i \gamma^5 + 16 \gamma^4 \Delta\omega_{21} + \Delta\omega_{21}(4(\Delta\omega_{21} - \Delta\omega_{23}) \Delta\omega_{23} + \Omega_{12}^2)^2 \right. \nonumber \\
 && + ( 4 (2 \Delta\omega_{21}^3 - 5 \Delta\omega_{21}^2 \Delta\omega_{23} + 2 \Delta\omega_{21} \Delta\omega_{23}^2 + \Delta\omega_{23}^3) - (\Delta\omega_{21} - 3 \Delta\omega_{23}) \Omega_{12}^2) \Omega_{23}^2 \nonumber \\
 && + ( - 2 \Delta\omega_{21} + 3 \Delta\omega_{23}) \Omega_{23}^4 + 8 i \gamma^3 ( 2 \Delta\omega_{21}^2 - 4 \Delta\omega_{21} \Delta\omega_{23} + 4 \Delta\omega_{23}^2 + \Omega_{12}^2 + \Omega_{23}^2) \nonumber \\
 && + 4 \gamma^2(  2 \Delta\omega_{21} (2 \Delta\omega_{21}^2 - 4 \Delta\omega_{21} \Delta\omega_{23} + 4 \Delta\omega_{23}^2+ \Omega_{12}^2) - (\Delta\omega_{21} - 2 \Delta\omega_{23}) \Omega_{23}^2   ) \nonumber \\
&& \left.  + i \gamma ( ( 4 (\Delta\omega_{21}-\Delta\omega_{23}) \Delta\omega_{23} + \Omega_{12}^2)^2 + 2 ( 6 \Delta\omega_{21}^2 - 14 \Delta\omega_{21} \Delta\omega_{23} + 8 \Delta\omega_{23}^2 + \Omega_{12}^2) \Omega_{23}^2 + \Omega_{23}^4   )         \right) \nonumber \\
{\rm NUM^V_{13}} &=& \Omega_{12} \Omega_{23} \left(  16 \gamma^4 - 24 i \gamma^3 (\Delta\omega_{21} -\Delta\omega_{23}) + 8 \Delta\omega_{21}^3 \Delta\omega_{23}  \right. \nonumber \\
&& - 16 \Delta\omega_{21}^2 \Delta\omega_{23}^2 + 8 \Delta\omega_{21} \Delta\omega_{23}^3 + 2 \Delta\omega_{21}^2 \Omega_{12}^2  \nonumber \\
&& + (2 \Delta\omega_{21} d23  - 4 \Delta\omega_{23}^2 + \Omega_{12}^2) \Omega_{12}^2 + 2 (-2 \Delta\omega_{21}^2 + \Delta\omega_{21} \Delta\omega_{23} + \Delta\omega_{23}^2+ \Omega_{12}^2) \Omega_{23}^2 + \Omega_{23}^4 \nonumber \\
&&  + 8 \gamma^2 (2 \Delta\omega_{21} \Delta\omega_{23} + \Omega_{12}^2 + \Omega_{23}^2) - 2 i \gamma (\Delta\omega_{21}-\Delta\omega_{23})( 4 (\Delta\omega_{21}^2 -\Delta\omega_{21} d23 + \Delta\omega_{23}^2) \nonumber \\
&& \left. + 3 (\Omega_{12}^2 + \Omega_{23}^2) ) \right) \nonumber \\
{\rm NUM^V_{23}} &=& \Omega_{23} \left( -16 i \gamma^5 + 16 \gamma^4 \Delta\omega_{23} + 16 \Delta\omega_{21}^4 \Delta\omega_{23} + 4 \Delta\omega_{21}^3 ( - 8 \Delta\omega_{23}^2 + \Omega_{12}^2)  \right. \nonumber \\
&& + 8 \Delta\omega_{21}^2 \Delta\omega_{23} (2 \Delta\omega_{23}^2 + \Omega_{12}^2 - \Omega_{23}^2)          \nonumber \\
&& + 3 \Delta\omega_{21} \Omega_{12}^2 (\Omega_{12}^2 + \Omega_{23}^2) - 8 i \gamma^3 (4 \Delta\omega_{21}^2 - 4 \Delta\omega_{21} \Delta\omega_{23} + 3 \Delta\omega_{23}^2 + \Omega_{12}^2 + \Omega_{23}^2) \nonumber \\
&& + \Delta\omega_{21} \Delta\omega_{23}^2 (-20 \Omega_{12}^2 + 8 \Omega_{23}^2) \nonumber \\
&& + 4 \gamma^2 (8 \Delta\omega_{21}^2 \Delta\omega_{23} - 8 \Delta\omega_{21} \Delta\omega_{23}^2 + 4 \Delta\omega_{23}^3 + 3 \Delta\omega_{21} \Omega_{12}^2 - \Delta\omega_{23} \Omega_{12}^2 + 2 \Delta\omega_{23} \Omega_{23}^2 ) \nonumber \\
&& + \Delta\omega_{23}( 8 \Delta\omega_{23}^2 \Omega_{12}^2 - 2 \Omega_{12}^4 - \Omega_{12}^2 \Omega_{23}^2 + \Omega_{23}^4) -i \gamma ( 16 \Delta\omega_{21}^4 - 32 \Delta\omega_{21}^3+12\Delta\omega_{23}^2 \Omega_{12}^2 \nonumber \\
&& \left. 8 \Delta\omega_{21}^2 ( 2 \Delta\omega_{23}^2 + 2 \Omega_{12}^2 - \Omega_{23}^2  ) + (\Omega_{12}^2 + \Omega_{23})^2 + 4 \Delta\omega_{21} \Delta\omega_{23} (-7 \Omega_{12}^2 + 2 \Omega_{23}^2) ) \right)  \nonumber
\end{eqnarray}  
where DEN is the same as for the Ladder-type solution give above in eq\,\ref{eq:DEN} but with $\Delta\omega_{12}$ replaced by $-\Delta\omega_{21}$.   It is noted
that the expressions for $\rho_{11}^{\rm V} $ and $\rho_{33}^{\rm V} $ are symmetric under the exchange $ 1 \leftrightarrow 3$ in the subscripts.  

The case where the initial population is entirely in
state $3$ can be calculated as the identify matrix minus the density matrices for initial population in states $1$ and $2$, 
or by inverting the $1$ and $3$ labels in the ladder-type DR expressions.  The general case, with thermal population in all three levels, can be written as the sum of these three
density matrices weighted by the fractional equilibrium population in each of the three states.

 \end{document}